\documentclass[11pt,a4paper]{article}
\usepackage[utf8]{inputenc}
\usepackage[T1]{fontenc}
\usepackage{lmodern}
\usepackage{microtype}
\usepackage[margin=1in]{geometry}
\usepackage{setspace}
\usepackage{graphicx}
\usepackage{booktabs}
\usepackage{amsmath,amssymb}
\usepackage{authblk}
\usepackage{xcolor}
\usepackage[hidelinks,colorlinks=true,linkcolor=blue,citecolor=blue,urlcolor=blue]{hyperref}
\usepackage[numbers,round,super,sort&compress]{natbib}
\usepackage{titlesec}
\usepackage{caption}
\usepackage{tabularx}
\usepackage[table]{xcolor}
\usepackage{color,soul}
\usepackage{ragged2e}
\usepackage{longtable}
\usepackage{booktabs}
\usepackage{array}
\usepackage[normalem]{ulem} % [normalem] prevents the package from changing \emph to underline
\DeclareUnicodeCharacter{2212}{\ensuremath{-}}
\usepackage{setspace}
\usepackage{lineno}

\newcolumntype{P}[1]{>{\RaggedRight\arraybackslash}p{#1}}

\titleformat{\section}{\normalfont\bfseries\large}{}{0pt}{}
\titleformat{\subsection}{\normalfont\bfseries\normalsize}{}{0pt}{}

\definecolor{boxteal}{HTML}{01696F}
\definecolor{boxbg}{HTML}{F2F4F4}
\newsavebox{\boxonecontent}

\title{\baselineskip=18pt \textbf{Bidirectional Wireless Communication for Weakly Coupled Implantable Brain-Computer Interfaces}}
 
\author[1,*]{Shreyas Sen}
\author[2]{Baibhab Chatterjee}
\author[1]{Gourab Barik}
\author[1]{Anirudh Roy}

\affil[1]{\baselineskip=14pt Elmore Family School of Electrical and Computer Engineering, Purdue University, West Lafayette, IN 47907, USA.}
\affil[2]{Department of Electrical and Computer Engineering, University of Florida, Gainesville, FL 32611, USA.}
\affil[*]{\href{mailto:shreyas@purdue.edu}{shreyas@purdue.edu}}
\date{}

\begin{document}
\maketitle
\baselineskip=14pt
%========================================================================================
\begin{abstract}
\baselineskip=14pt

Implantable brain-computer interfaces (BCIs) promise transformative societal impact, from restoring lost motor, sensory, and speech function in patients with paralysis, stroke, sclerosis, and sensory deficits to serving as a high-bandwidth conduit between human cognition and machine intelligence. Realizing these visions requires moving from laboratory prototypes to chronic clinical systems that support thousands to millions of electrodes, several millimeters to centimeters deep beneath the brain surface, under strict heating and size limits. Almost every clinically relevant implant therefore operates in a \emph{weakly coupled} regime, with coupling coefficients of $10^{-3}$ to $10^{-1}$ across centimetres of lossy tissue. In this regime, the wireless channel sets the limits of power-transfer efficiency, communication bandwidth, and energy per bit. We review the recent progress of bidirectional wireless-links across inductive, mid-field, RF, ultrasonic, magnetoelectric, optical, UWB, and electro-quasistatic modalities, and benchmark them against the clinical axes of depth, size, and data rate. Within a $\sim$10\,mW communication budget set by the $\sim$1\,$^{\circ}$C tissue-heating ceiling, Narrowband high-Q links are well suited for power transfer and low-speed data, but at 1-10\,nJ$/$b cannot reach the $>$10\,Mbps to 10's of Gbps uplinks that thousand to million-channel interfaces demand, even with aggressive on-implant compression. This calls for sub-10\,pJ$/$b and ultimately sub-1\,pJ$/$b wireless-links, where wideband techniques, such as ultra-wideband (UWB) and brain-channel communication (BCC) are suitable. We close with a quantitative framework for analyzing such wireless-links and a co-design roadmap across electromagnetics, circuits, packaging, security, and regulation, toward secure, networked, million-channel brain interfaces.

\end{abstract}
%========================================================================================
\section*{Key points}
\begin{itemize}

  \item Weak coupling ($10^{-3}$ to $10^{-1}$) is the primary modality  for small, deep BCIs, making wireless power and data transfer the primary bottleneck, demanding joint co-design of physics modality, transducer, and circuits.

  \item A $\sim$1\,$^{\circ}$C thermal ceiling caps the to tens of mW, and streaming 10k to 1M channels ($\sim$480\,Mbps to $>$10\,Gbps) calls for sub-10\,pJ$/$b wireless-links today and sub-1\,pJ$/$b in future.

\item Both wireless power-transfer efficiency (critical for downlink) and communication energy-efficiency (critical for uplink) are governed primarily by tissue loss, SAR, available bandwidth, and the chosen physics modality.

\item Resonant narrowband links (Inductive, ME) are attractive for high-PTE downlink, while wideband modalities (UWB, BCC) are emerging as the path to high-rate uplink, with recent cortical demonstrations achieving $\sim$100$\times$ of today's nominal 1Mbps links.

\item Aggressive on-implant lossy compression ($10^{2}$ to $10^{4}\times$, e.g. the Neuralink Challenge at $\sim$200$\times$) is a workaround for energy-scarce links, not a destination, since lossless limits sit near $\sim$10$\times$; reaching a near-raw uplink will require new innovations that drive whole-system energy from $>$500\,pJ$/$b down toward 10 to 50\,pJ$/$b.

\end{itemize}
%========================================================================================
\section*{Introduction}

Brain-computer interfaces (BCIs) record electrical neural activity for external processing or stimulate targeted regions of the brain through patterned electrical, optical, magnetic, or acoustic signals.  Two decades of intracortical recording in humans have established that even a few hundred well-isolated cortical units can support typing\cite{ref1,ref2,ref3}, speech decoding\cite{ref4,ref5,ref6,ref7}, and prosthetic limb control\cite{ref2,ref136,ref137} in patients with paralysis, stroke, sclerosis, and sensory deficits. Larger cortical-surface arrays extend these capabilities while avoiding cortical penetration, and endovascular arrays reach motor cortex without craniotomy at all, opening BCIs to patient populations\cite{ref8,ref9,ref13}.
Between 2021 and 2026 the field reached an inflection point, marked by the first at-home use of a wireless intracortical interface\cite{ref11}, multi-hundred-channel commercial implants entering human trials, the first fully internalised speech neuroprosthesis decoding inner speech\cite{ref7}, and millimetre-scale implants powered ultrasonically or magneto-electrically for chronic stimulation\cite{ref23,ref24,ref25,ref26,ref27,ref28,ref29,ref30,ref112}. Beyond the clinical applications, the same technology base is positioned to serve as a high-bandwidth conduit between human cognition and machine intelligence.

A unifying technical challenge underlying these developments the \emph{wireless link}. Chronic human use calls for the implant to be battery-free, drawing its operating energy from an external source through the same tissue volume that returns its data. The implant must acquire signals from $10^{2}$ to $10^{6}$ electrodes, digitize and partially process them, convey data,  receive commands,  deliver any stimulation current locally, and ideally provide low-latency bidirectional control, through skin, skull, dura, and brain tissue \emph{without a wired tether}, to avoid challenges related to infection, tissue-damage, micro-motion due to wire strain, as well as mechanical failure, while improving scalability and patient mobility\cite{ref16}. All functionalities must fit within tens of milliwatt power budget\cite{ref150,ref67,ref68}, in volumes from $10^{-3}$\,mm$^{3}$ to a few cm$^{3}$, at depths from 3\,mm subdural to $\sim$50\,mm in the deep brain, with chronic operation expected over more than ten years. 

The dominant physical constraint on the link itself is \emph{weak coupling}\cite{ref19,ref20,ref21,ref25,ref90,ref110}. The implant antenna or transducer is small, the external coil or array has to power and communicate through a lossy tissue medium, and most physical links deliver a small fraction of the transmitted energy. Practical coupling coefficients fall between $10^{-3}$ and $10^{-1}$, making the channel itself the dominant bottleneck and demanding joint co-design of physics modality, transducer geometry, and circuits.

Another key constraint is \textit{thermal}. A $\sim$1\,$^{\circ}$C tissue-heating ceiling caps total implant power near 70\,mW and, after safety margins and on-chip overhead, the wireless link itself near 10\,mW. Within this envelope, streaming 10,000 to one million channels corresponds to $\sim$500\,Mbps at the low end and exceeds 10\,Gbps at the high end. Conventional narrowband links operating at 1 to 10\,nJ$/$b cannot close this gap, forcing link energies below 10\,pJ$/$b in the near term and toward 1\,pJ$/$b in the longer term.
 
In this Review, we cover the recent progress of various wireless-link modalities such as inductive, mid-field,  radiofrequency, ultrasonic, magnetoelectric (ME), optical, ultrawideband (UWB), and electro-quasistatic and mid-field BCC, highlighting that each occupy a distinct point in this design space, and the right choice depends on the depth, size, and data-rate requirements of the target application. A coherent treatment must therefore begin from clinical \emph{application needs}, highlight the \emph{physics of the wireless-links} matching variable needs supported by academic literature,  followed by examination of the \emph{commercial} state of the art, and only then \textit{compare} modalities through a common framework to highlight what's possible today and where the biggest opportunity exists.

%In this Review, we take weak coupling as the organising principle and develop the argument in five parts. Section 1 frames the wireless layer from the clinical end, defining what downlinks must carry (power, stimulation commands, configuration, firmware) and what uplinks must convey (broadband spikes, LFP, ECoG and $\mu$ECoG streams, closed-loop biomarkers), with explicit quantitative targets for rate, latency, and bit-error rate. Section 2 formalises the weakly coupled regime through link budgets, channel capacity, the Chu limit on electrically small antennas, and the thermal and regulatory safety envelope. Section 3 surveys the commercial landscape, organising fielded and trial-stage systems by clinical category. Section 4 develops a quantitative modality-by-modality comparison of downlink and uplink channel capacity, benchmarking candidate modalities on a common energy-per-bit and data-rate plane. Section 5 closes with forward-looking research directions: co-design across electromagnetics, transducers, circuits, packaging, and regulation; on-implant compression and inference; post-quantum-secure links and physical-layer security; networks of implants; and more speculative but plausible directions including programmable metasurfaces, reservoir-style analog computation, and biological transduction.
%========================================================================================
\section*{1. Application Needs}

%========================================================================================
% Fig 1 (single figure, six panels a--f)
\begin{figure}[htbp]
\centering
\includegraphics[width=0.95\textwidth]{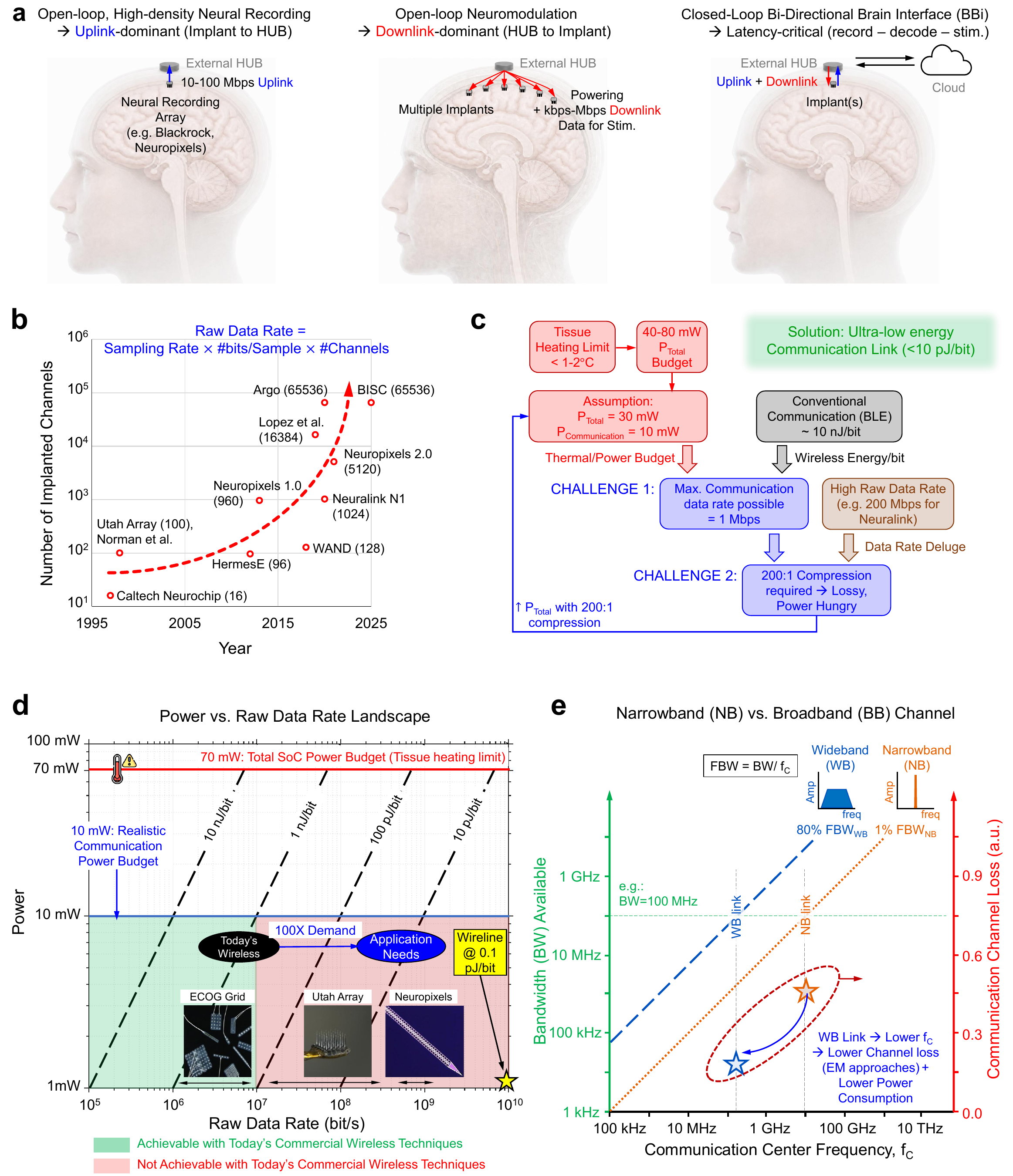}
\caption{\textbf{Application classes of weakly-coupled wireless implantable BCI and relevant design constraints.}
  \textbf{a}, Three application classes of today's implantable wireless BCI: Open-loop high-density neural recording, Open-loop neuromodulation, and Closed-loop Bidirectional Brain Interface (BBI). \textbf{b}, current trends showing an exponential increase in the number of channels for neural interfaces. \textbf{c}, Uplink data rate limitations due to conventional high-energy communication, requiring lossy and power hungry compression at the node, which in turn increases system-level power consumption. More efficient communication can alleviate the bottleneck. 
  \textbf{d}, Uplink power-vs-raw-data-rate landscape: today's devices with $\sim$10,000-channels produce $\sim$1000\,Mbps or more raw data.  With a $\sim$10\,mW uplink power budget, sub-10\,pJ/b links are required, with a sub-1\,pJ/b aspiration for million-channel raw streaming.
  \textbf{e}, Narrowband, high-$Q$ resonant links for power and commands (FBW $\sim$0.1 to 1\,\%) versus wideband, low-$Q$ links for high-rate data (FBW $\sim$80\,\%). To achieve the same bandwidth, wideband links require lower center frequencies, leading to lower link loss and lower power consumption.}
\label{fig:intro}
\end{figure}

% \subsection*{Three application classes}
\subsection*{Implantable BCI application classes}

We start this Review of wireless techniques in brain-implants not from the links themselves but from the clinical tasks they must serve. Fielded and emerging systems fall into three application classes, as shown in Fig.~\ref{fig:intro}a: \textbf{~open-loop high-density neural recording} (uplink-dominated); \textbf{~open-loop neuromodulation} (downlink-dominated); and \textbf{~closed-loop Bidirectional Brain Interface (BBI)}, requiring recording and stimulating inside one sub-10\,ms local loop (latency-limited). The amount of compute that each device can host on the implant versus a body-worn hub, a phone, or the cloud, as well as the application complexity and cost determines which class is realizable in practice, and the quantitative requirements and trade-offs in Fig.~\ref{fig:intro}b--e drive the modality and architecture choices analyzed in the rest of this Review. The three classes are described below, each closing with the uplink/downlink constraints in terms of energy, bandwidth, and data rates, along with the compute-tier assignment that makes it deployable.

\subsection*{(a) Open-loop high-density neural recording}
    \label{sec:Application_classes_A}
    %\sout{
    Many of today's high-density intracortical recording arrays are using 100 to $\geq$1024-channel Utah-array and Neuropixels-class probes to produce multi-Mbps throughputs. Neuralink's N1-class platforms (1024 channels), now in human trials\cite{ref10,ref11,ref12,ref14,ref15,ref41}, produce $\sim$200 Mbps raw data, to be transferred out of the brain at only tens of milliwatts of power. Single-channel neural data may be generated at only $\sim$1-20\,kSps, but with 10--16 bits analog to digital conversion (ADC) per channel, the data rate scales rapidly with larger arrays. As an example, $\sim$100\,Mbps data rates were required for the 5376-channel $\mu$ECoG array demonstrated in a head-mounted device in 2024\cite{ref89}. Data rates are projected to reach 10\,Gbps (e.g. 16b, 30ksps, 20k channels or 10b, 1ksps, 1M channels) for high resolution fast sampling, high-channel platforms in the next decade\cite{ref141}. Such exponential increase in number of channels (Fig. \ref{fig:intro}b) and data rates exceeds the current capabilities of achievable transmission rates, within the implant's power and thermal budget, making wireless communication one of the primary bottlenecks in terms of the system power budget.
    
    Beyond spiking activities, the uplink also carries local field potentials (LFP) at 1 to 500\,Hz per channel for state monitoring, ECoG and $\mu$ECoG signals at hundreds of Hz to a few kHz, and slow biomarkers (impedance, temperature, electrochemistry) used for health monitoring and for closed-loop control of stimulation\cite{ref122,ref147}. Bit-error-rate (BER) targets are stringent: raw spike payloads tolerate BER in the 10$^{-6}$ to 10$^{-4}$ range because errors translate to added decoder noise, while compressed or quantized streams demand a practical floor of $\sim$10$^{-9}$, motivating per-packet forward error correction (FEC) even at 10 to 15\,\% rate overhead\cite{ref41,ref124}. Such overheads increase the communicable payload, unless aggressive compression is used, which itself is power hungry.

    The link-energy specification for this class follows from tissue thermal limits. Sustained heating of neural tissue must stay below $\sim$1\,$^\circ$C to avoid neuronal injury, capping whole-implant power at $\sim$70\,mW for a few-cm$^3$ device\cite{ref23,ref24,ref25}. A factor-of-two safety margin against worst-case perfusion reduces the design budget to $\sim$35\,mW, and the wireless transceiver typically claims about one-third of that envelope (the rest going to the front-end amplifiers, ADCs, on-implant compute, and stimulation drivers), leaving $\sim$10\,mW for the link: the horizontal ceiling drawn in Fig.~\ref{fig:intro}d. A 10\,000-channel implant sampling at 4\,kSps and 12 bits produces $\sim$480\,Mbps of raw broadband neural data; meeting that rate below the 10\,mW ceiling requires sub-10\,pJ/b links, and the million-channel future ($\sim$10\,Gbps) demands sub-1\,pJ/b operation. Today's deployed wireless BCI links sit one to three orders of magnitude above this threshold, in the $\sim$100\,pJ/b to 10\,nJ/b range\cite{ref41,ref89,ref141}, whereas wirline links are often in the range of  $\sim$0.1\,pJ/b.

    The prevailing solution is to reduce the amount of data that must be transmitted out of the implant: on-die threshold crossing, spike sorting, compressive sensing, learned codecs, and small embedded neural networks can achieve 10$^{2}$ to 10$^{4}\times$ data reduction\cite{ref35,ref36,ref44,ref54,ref123,ref124,ref125,ref126,ref76}. The Neuralink Compression Challenge aimed for a $\sim$200$\times$ lossless compression (raw $\sim$200\,Mbps to $\sim$1\,Mbps wireless egress)\cite{ref41} to be compatible with their N1-device's throughput. Subsequent analysis has shown that information-theoretic limits on \emph{lossless} compression of broadband neural data sit closer to $<$10$\times$ due to the Johnson-Nyquist limits of background noise in the neural data\cite{neuralink_challenge_soln}.
    %anything beyond that ratio must be lossy and amounts to a hand-chosen feature commitment, such as spike count, threshold crossing, or a learned codec trained on yesterday's labels. 
    Fig. \ref{fig:intro}c-d shows that for a realistic communication power budget of $\sim$10 mW, and with the $\sim$10 nJ/b energy efficiency of conventional wireless (\emph{e.g.} BLE, as implemented in Neuralink's N1), the data throughput is limited to only $\sim$1 Mbps, necessitating the lossy on-implant compression. % is the workhorse today because link energy is scarce, but
    This invariably limits which features ever leave the brain and what is discoverable with the limited data: sharp-wave and fast ripples, theta/gamma and beta/gamma cross-frequency coupling, spike-waveform features that index cell type and dendritic state, sub-threshold fluctuations on high-density CMOS arrays, and the population covariance that underwrites manifold decoding are precisely the phenomena that thresholding and per-channel quantization discard\cite{ref147}. Reaching sub-10\,pJ/b is therefore the engineering \emph{need} that lets implants deliver clinically lossless or modestly compressed streams. On the other hand, reaching sub-1\,pJ/b is the longer-term \emph{aspiration} that opens a raw scientific uplink for foundation-model-class analysis (\emph{e.g.}, POYO, BrainBERT, and population-scale emerging broadband pipelines)\cite{ref141}. The same electrode array can therefore sit in one of three regimes set by the link's energy-per-bit: a \emph{compressed clinical regime} delivering $\leq$10\,Mbps at $\sim$100\,pJ/b; a \emph{near-lossless regime} that compresses the data by $\sim$10$\times$ and opens at sub-10\,pJ/b; and a \emph{raw scientific uplink} that streams the full electrode rate off-implant with no on-die feature commitment, thermally comfortable at sub-1\,pJ/b (at 1\,pJ/b, 480\,Mbps costs only 0.48\,mW). Such energy efficiencies are only possible in broadband/wideband systems, and hence, the recording uplink must sit on the broadband, low-$Q$ regime of Fig.~\ref{fig:intro}e, and pushes the modality choice toward UWB or BCC techniques (analyzed in Section 4).

\subsection*{(b) Open-loop neuromodulation}
    \label{sec:Application_classes_B}
    
    The open-loop neuromodulation class is fundamentally downlink-dominant: unlike neural recording systems, where uplink data transmission dominates the wireless budget, an external transmitter delivers multi-milliwatt power and kbps command traffic to one or many distributed stimulators for neuromodulation.
    Clinical platforms such as cochlear implants, deep-brain stimulation (DBS), spinal-cord stimulation (SCS), vagus-nerve stimulation (VNS), hypoglossal-nerve stimulation (HNS), and responsive neurostimulation (RNS) such as NeuroPace's epilepsy treatment device\cite{ref71,ref72,ref73,ref74,ref84,ref101,ref122}, all rely on the chronic delivery of \textbf{milliwatt-scale power} to support charge-balanced bi-phasic pulses \cite{ref71,ref72,ref73,ref74,ref101} for therapeutic stimulation. For example, Cochlear implants pioneered the multi-milliwatt transcutaneous inductive link more than three decades ago\cite{ref71,ref72,ref73}, and that architecture remains the template for almost every fielded neuromodulator, including the Medtronic Activa and Percept DBS families, Boston Scientific's Vercise platform, Inspire's hypoglossal-nerve stimulator, and NeuroPace's RNS responsive epilepsy system\cite{ref84,ref122}.    
    %This is the regime that has carried cochlear implants, deep-brain stimulation (DBS), spinal-cord stimulation (SCS), vagus-nerve stimulation (VNS), hypoglossal-nerve stimulation (HNS), responsive epilepsy systems such as NeuroPace RNS, and emerging cortical neuromodulators into routine clinical use\cite{ref71,ref72,ref73,ref74,ref84,ref101,ref122}. The downlink carries three classes of traffic. The first and most demanding is \emph{electrical power}: stimulators for cochlear processing, DBS, SCS, VNS, HNS, and cortical neuromodulators need 1 to 10\,mW or more delivered chronically to support charge-balanced biphasic pulses at therapeutic amplitudes\cite{ref71,ref72,ref73,ref74,ref101}, whereas pure recorders for spikes, LFP, ECoG, and $\mu$ECoG operate at sub-mW total dissipation set by the front-end and transceiver\cite{ref35,ref36,ref54,ref95}. Cochlear implants pioneered the multi-milliwatt transcutaneous inductive link more than three decades ago\cite{ref71,ref72,ref73}, and that architecture remains the template for almost every fielded neuromodulator, including the Medtronic Activa and Percept DBS families, Boston Scientific's Vercise platform, Inspire's hypoglossal-nerve stimulator, and NeuroPace's RNS responsive epilepsy system\cite{ref84,ref122}.
    
    Communication represents a secondary requirement and is largely limited to stimulation commands (\emph{e.g.}, amplitude, pulse width, frequency, channel map, biphasic balance), device configurations (\emph{e.g.}, recording-channel maps, gain and filter coefficients, codec selections, security features), and occasional over-the-air (OTA) firmware updates, all of which may have data integrity challenges, but do not require high-speed communication. Consequently, the required data rates are typically modest, ranging from kbps for parameter updates to at most Mbps for high-channel-count stimulators \cite{ref27,ref28,ref30,ref103}. For designs with shared  single inductive or mid-field carrier, the payload data may be simultaneously transmitted along with power by using reserve sidebands or duty-cycled windows\cite{ref93,ref94,ref95,ref96,ref120};  or by modulating various properties of the powering pulse, as shown in ultrasonic and magnetoelectric (ME) systems\cite{ref25,ref28,ref29,ref30,ref105} .

    The downlink-dominant applications naturally favor narrowband, high-$Q$ wireless links. Maximizing power-transfer efficiency (PTE) is substantially more important than maximizing data throughput, making resonant inductive, mid-field RF, ultrasonic, and magnetoelectric links particularly attractive for downlink operation. As a result, most practical neuromodulation systems employ a dedicated low-bandwidth channel optimized for efficient power and command delivery, rather than for high-rate telemetry \cite{ref27,ref28,ref30,ref93,ref94,ref95,ref96,ref103,ref120}. %A wireless link's operating point is fixed by its fractional bandwidth FBW $= BW/f_{c}$, where $f_{c}$ is the center frequency used. BLE (FBW $\sim$0.1\,\%) and Wi-Fi-class carriers (FBW $\sim$1\,\%), as well as the resonant inductive and mid-field schemes used by commercial devices sit at FBW $\ll$ 1\,\%.
    This fundamental asymmetry between power delivery and data transmission is one of the primary reasons why future bidirectional brain interfaces are likely to employ different modalities for the downlink and uplink paths, a concept that we shall revisit in Section 4.

\begin{figure}[t]
\centering
\includegraphics[width=1\textwidth]{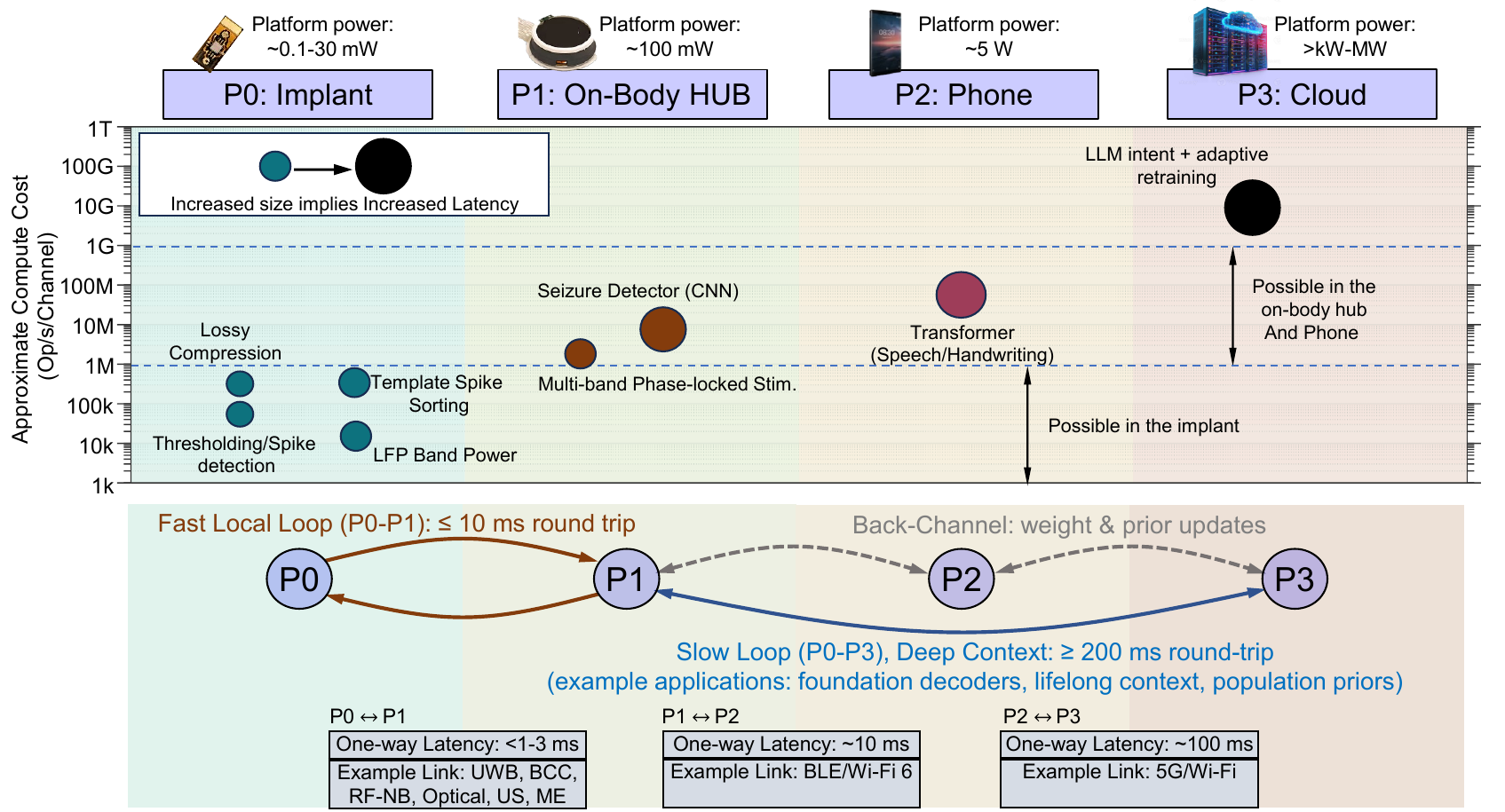}
\caption{Four-tier compute hierarchy: implant (P0), body-worn hub (P1), phone (or similar device - P2), cloud (P3) with a fast P0$\leftrightarrow$P1 local loop ($\leq$10\,ms) and a slow P0\,$\rightarrow$\,P3 deep loop. Latencies, powers and exemplar computations are shown for each tier, showing which computations are possible in which tiers.}
\label{fig:latency}
\end{figure}

\subsection*{(c) Closed-loop Bidirectional Brain Interface (BBI)}
    \label{sec:Application_classes_C}

    The closed-loop BBI combines the requirements of neural recording and neuromodulation within a single system, that records and stimulates within a sub-10\,ms local-loop latency so that motor, speech, or therapeutic feedback feels natural to the user. It therefore inherits both the high-throughput uplink, and the resonant,  high-PTE downlink on a single shared channel, with two new constraints as described next. The first is end-to-end latency. Implants engaged in motor or speech BCI control must meet closed-loop latency below $\sim$50\,ms to feel natural, and ideally below $\sim$10\,ms\cite{ref45}, ruling out store-and-forward strategies that batch large blocks of compressed data even when batching would help link energy. 
    
    %The second is compute placement: realizing a sub-10\,ms loop requires placing each computation on the proper tier of Fig.~\ref{fig:latency}. The four tiers carry very different power and latency budgets: P0 denotes the implant ($\sim$mW, $<$1 to 5\,ms one-way), P1 is the body-worn hub ($\sim$100\,mW, 5 to 10\,ms), P2 could be a on-body device such as a phone ($\sim$5\,W, 30 to 100\,ms), and P3 represnets cloud (kW to MW, $>$100\,ms). On-implant compute is bounded by $\sim$10\,mW thermal envelope of class (a), so there is a threshold in compute cost beyond which a task must be offloaded to the next tier.

    %\textcolor{orange}{
    %Closed-loop bidirectional brain interfaces (BBIs) combine the requirements of neural recording and neuromodulation within a single system. In addition to supporting both a high-throughput uplink and a powered downlink, these systems introduce a third and often more stringent requirement: low end-to-end latency. Neural activity must be recorded, interpreted, and translated into stimulation rapidly enough to preserve natural motor, sensory, or therapeutic feedback, typically within a sub-10,ms to tens-of-milliseconds timescale \cite{ref45}.

    The second is compute placement: meeting these latency targets requires careful placement of computation across multiple processing tiers, as shown in Fig.~\ref{fig:latency}. Simple and latency-critical operations can be performed directly on the implant, whereas computationally intensive tasks such as neural decoding, adaptive stimulation, and model updating are increasingly offloaded to a body-worn hub, personal device, or cloud infrastructure. The resulting architecture forms a hierarchy spanning the implant (P0), body-worn hub (P1), a phone-like device (P2), and cloud (P3), each offering progressively greater computational capability at the cost of increased latency and power consumption. Example applications, such as compression, CNN-based seizure detection and template spike sorting, along with their  associated compute cost (operations per channel per second) and the proper hierarchal location are also outlined in the figure. This hierarchy introduces a fundamental trade-off between communication and computation. Aggressive on-implant processing can reduce wireless bandwidth requirements but increases local power consumption and latency. Conversely, sufficiently energy-efficient communication enables more neural data to be transmitted off-implant, where substantially greater computational resources are available, at a latency cost. As communication energy approaches the sub-10\,pJ/b regime, raw or lightly processed neural streams become increasingly practical, reducing the need for aggressive feature extraction while enabling richer decoding and adaptive stimulation within the same thermal and latency limits.
    
    Closed-loop BBIs therefore represent the most demanding application class considered in this Review, requiring the simultaneous optimization of power delivery, communication efficiency, latency, and compute placement. The energy-per-bit, bandwidth, and compute-tier constraints introduced across the three application classes set the agenda for the channel-capacity, modality, and architecture analyses of Section 2 to Section 4.

\section*{2. Physics of Wireless Technique employed in Brain Implants}

\subsection*{Weakly coupled wireless interaction with a brain implant}

\begin{figure}[t]
   \centering
    \includegraphics[width=1\textwidth]{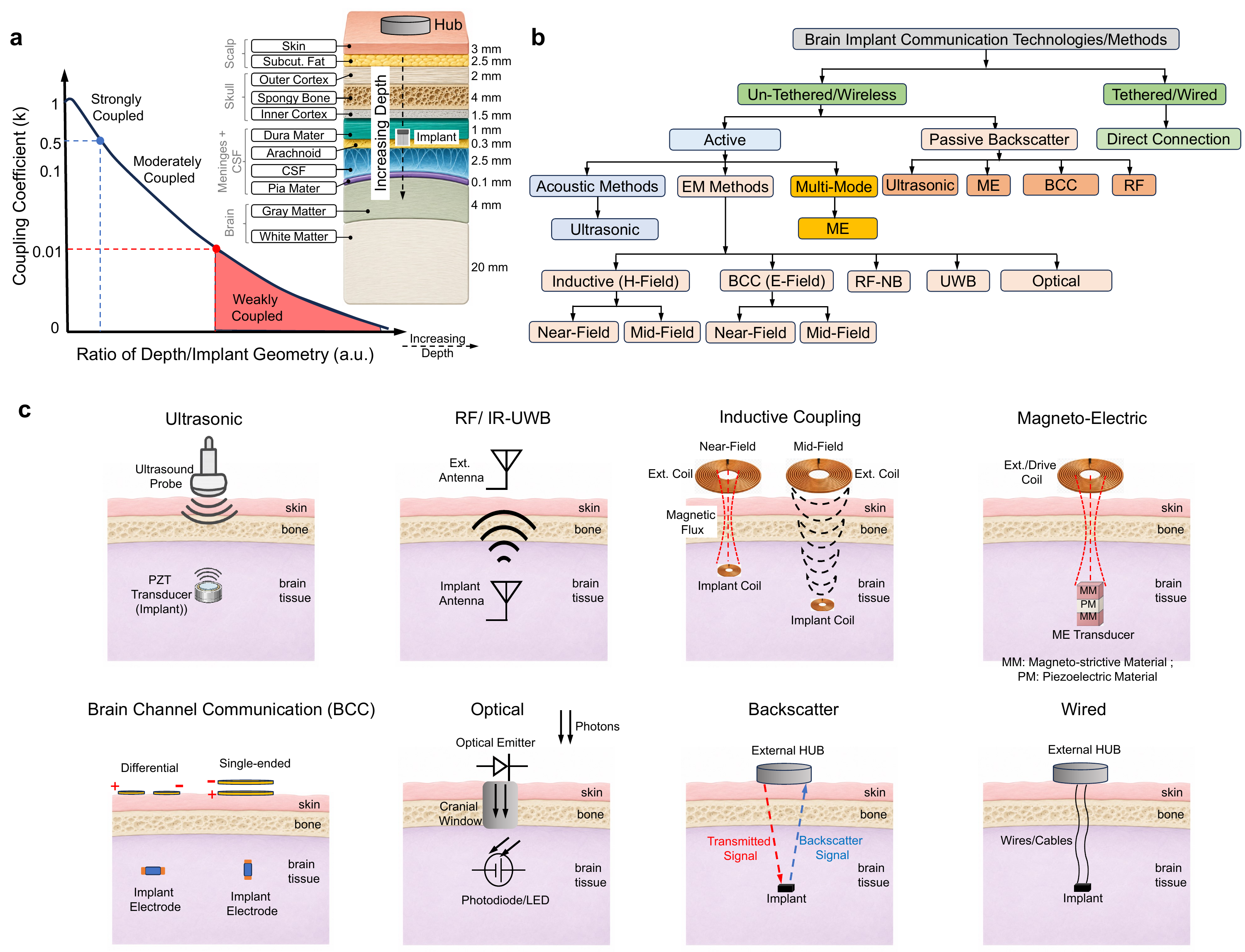}
     \caption{\textbf{Brain Implant Communication Technologies. a,} weakly-coupled region definition. \textbf{b,} available techniques (tethered/untethered) for communication with a brain implant. \textbf{c,} description of popular techniques: ultrasonic, RF, inductive coupling, magneto-electric, brain channel communication, optical, backscatter and wired technologies in the current context.}
    \label{fig:tech}
    \vspace*{-0in}
\end{figure}

For any two electromagnetic (EM) or acoustic transducer/resonator/antenna/couplers separated by a medium, the coupling coefficient \emph{k} represents the fraction of mutual energy exchanged per cycle. Practical brain implants almost always operate at \emph{k} = 10$^{-3}$ to 10$^{-1}$ range.
%: the implant coil/antenna/transducer is much smaller than the external one, \textcolor{red}{not true: the separation exceeds the implant's largest dimension by an order of magnitude}, and tissue loss attenuates whatever fields do cross the boundary\cite{ref20,ref21,ref22,ref70,ref71,ref90}.
In this regime, the implant aperture is electrically small and separated from the external source by multiple implant diameters in the lossy tissue\cite{ref20,ref21,ref22,ref70,ref71,ref90}.
%so only a fraction $k^2 \approx$ 10$^{-6}$ to 10$^{-2}$ of the launched energy can be recovered per cycle\cite{ref20,ref21,ref22,ref70,ref71,ref90}.
This is shown in Fig. \ref{fig:tech}a, as a qualitative function of the ratio of depth to implant geometry. The wireless techniques that are used to communicate between implants and the external world can be categorized into (1) active and (2) passive (backscatter), as shown in Fig. \ref{fig:tech}b. Among the available wireless techniques, near-field inductive coupling (NIC), radio-frequency (RF), optical, ultrasonic (US), body/brain-channel communication, magnetoelectric (ME), and ultra-wideband (UWB) techniques are being widely explored due to the specific advantages that each modality offers. For any of the EM/acoustic propagation modalities, three independent regimes coexist depending on geometry and frequency: (i) the \emph{quasi-static near field} close to the source, where electric/magnetic fields decay follows 1/r$^{3}$ trends ($r$ = distance from source); (ii) the \emph{mid-field}, %where electrically large external sources sculpt evanescent components that propagate constructively to the implant\cite{ref20,ref21,ref22};
where electrically large external sources create evanescent components that focus energy at certain depths, improving coupling relative to purely quasi‑static links\cite{ref20,ref21,ref22}; and (iii) the \emph{far field}, used by MedRadio (402--405 MHz) and ISM-band (915 MHz, 2.45 GHz) telemetry\cite{ref69,ref70}. Fig. \ref{fig:tech}c shows the different active modes of wireless telemetry (Ultrasonic, inductive, RF/UWB, optical, magneto-electric, and brain-channel communication), along with passive backscatter and wired modes. For the uplink, the implant acts as the transmitter for the particular modality/field, while an external device (hub) receives the signal specific to each modality after it has propagated through the tissue.

%Sections 2 and 4 therefore treat weak coupling not as a qualitative label but as a quantitative operating point where achievable data rate is eventually set by a function of coupling efficiency (which sets the SNR) and usable bandwidth under safety limits. The capacity calculations in Section 4 explicitly anchor NIC and ME in the quasi‑static regime, US in the acoustic Fresnel/far‑field regime, and UWB/BCC in radiating or EQS regimes, so the taxonomy introduced here maps directly onto the modality‑specific models in Table \ref{tab:combined}.

US and ME modalities are mechanical analogues of the EM regimes, exploiting piezoelectric\cite{ref23,ref24,ref25,ref26,ref27,ref105,ref106,ref107,ref108,ref109} or magnetostrictive\cite{ref28,ref29,ref30,ref111,ref112} transduction,  respectively. Electro-quasistatic (EQS) human-body-communication/body-channel communication generally operate below 30-50 MHz where the dimension of the body-channel is much smaller than the wavelength, resulting in EQS fields within and around the body \cite{ref45,ref46,ref47,ref48,ref49,ref50,ref52,ref57,ref144}. For the brain, due to the reduced dimensions of the medium (as compared to the whole body), the EQS range extends to $\sim$ 100 MHz \cite{11509547}.
%At these frequencies, the  \emph{termination impedance} (either single-ended or differential, based on the signal coupling technique) becomes as important as the geometry of the implant electrode itself\cite{ref47,ref48}. Due to the small device size to wavelength ratio, this mode is conceptually equivalent to the near-field EM, with the additional constraint of termination impedances. %although the channel loss is significantly affected by termination (high impedance vs. low impedance).
For the rest of the paper, we shall refer to this modality as BCC, which interchangeably represents body-channel communication and brain-channel communication. Interestingly, one modality can outperform the others in terms of either channel loss/signal to noise ratio (SNR), safety or channel capacity at certain frequencies and geometries, but not across all frequencies and geometries . The result is a five-axis design space spanning (1) frequency, (2) coupling structure, (3) transducer physics, (4) safety and biocompatibility, and (5) channel termination.

\subsection*{Downlink modalities}

As explained in Section 2, downlink modalities with larger PTE (\emph{i.e.}, the ratio of received to transmitted power) is preferred for wireless powering. This is generally achieved with high-$Q$ resonances, which invariably limits the bandwidth and the data rate in low-to-medium (kbps-Mbps) range. The popular downlink modalities are described below.

\emph{Inductive coupling} in the kHz to tens-of-MHz range remains the dominant downlink for large, multi-milliwatt implants\cite{ref19,ref73,ref74,ref75,ref76,ref77,ref90,ref93,ref95,ref101}. Recent advances have pushed performance closer to fundamental limits through (i) high-\emph{Q} litz or printed coils with ferrite backing\cite{ref73,ref74}; (ii) class-E and class-D/E$^{2}$ primary drivers that exceed 90\% driver efficiency\cite{ref101}; and (iii) closed-loop coil-impedance tuning that absorbs coupling variation from head motion\cite{ref93,ref95}.
%and (iv) head-stages that distribute power uniformly over a 3-D volume for behavioral experiments\cite{ref75}.
PTE between 5\% and 60\% is achievable at depth (\emph{d}) $\leq$ 5 mm, but as \emph{d} grows beyond 15 mm and the implant coil diameter shrinks below 3 mm, PTE collapses below 0.1\%\cite{ref76,ref77,ref90}.
%For inductive links, fitted PTE–depth relations (e.g., PTE$_{ind}$(\emph{d}) $\propto$ \emph{d}$^{−3}$ for a fixed mm$^3$ receiver) quantify how rapidly efficiency collapses with depth, and show that tissue contributes $<$2 dB of loss at 30 mm compared to $>$10 dB from geometry (see Table \ref{tab:combined}, Section 4). 
Additionally, misalignment of the transmit and receive coils often reduces the PTE during inductive coupling.

\emph{Mid-field RF} deliberately exploits the GHz tissue--wavelength match: a patterned external source synthesizes evanescent components in the tissue that constructively reform around the deep implant\cite{ref20,ref21,ref22,ref90}. Prior demonstrations include a 2-mm cardiac stimulator at 5-cm depth\cite{ref20}, microwave-powered neuromodulators for the brainstem\cite{ref76}, and a mid-field-powered chronic recording implant in freely behaving primates\cite{ref150}.
%True far-field telemetry in MedRadio (402--405 MHz)\cite{ref64} and ISM bands is reserved for low-power \emph{data} rather than power, with effective isotropic radiated power (EIRP) capped at 25 $\mu$W$^{69}$,   adequate for $\leq$500 kbps command links but not for clinical-rate neural data\cite{ref41}.

\emph{Ultrasonic power transfer} bypasses the electromagnetic SAR ceiling because the regulatory limits leave more headroom for sub-10-MHz acoustic intensity\cite{ref23,ref24,ref25,ref26,ref27,ref105,ref106,ref107,ref108,ref109}. Berkeley's neural dust\cite{ref23,ref24}, and subsequent generations have demonstrated bidirectional telemetry\cite{ref25,ref27,ref106,ref107,ref108},
%and extended the concepts to oxygen\cite{ref106}, temperature\cite{ref107}, and pressure\cite{ref108} monitoring, 
with the smallest demonstrated end-to-end ultrasonic mote occupying only 0.065 mm$^{3}$ volume\cite{ref107}.
%Multi-mote networks have addressed crosstalk through frequency- and time-domain multiplexing\cite{ref26} and through adaptive ultrasonic powering that compensates for tissue motion\cite{ref109}.
The primary bottleneck for practical use of ultrasonic systems in the brain is the skull, which introduces $\geq$15 dB loss and severe phase aberrations at 1--10 MHz; RF-to-ultrasound relays at the skin/just below the skull overcome this for shallow targets\cite{ref79}.

\emph{Magnetoelectric} (ME) composites convert a kHz-range magnetic field into voltage with or without antenna resonance, allowing mm-scale implants to operate at depths and through skull regions where neither inductive nor mid-field RF performs well\cite{ref28,ref29,ref30,ref111,ref112}. MagNI\cite{ref28} and ME-BIT\cite{ref30} demonstrated chronic stimulation in rodent and pig models using 100--300 kHz fields below ICNIRP limits, with 8.2-mm$^{3}$ ME-powered closed-loop neurostimulators\cite{ref111}. However, the external device size is usually large to support larger magnetic fields.
%ME is particularly attractive for free-floating multi-implant networks because a single external coil can ccouple uniform magneto-quasistatic fields in the tissue, while each implant draws what it needs\cite{ref32}. 

\emph{Optical/NIR photovoltaic} powering is useful for small depths ($<$ 1 cm), due to scattering challenges. External LEDs (Light-Emitting-Diodes) or VCSELs (Vertical-Cavity Surface-Emitting Laser) supply 0.1--1 mW at conversion efficiencies of $\sim$5\%, and have powered optogenetic and photometric implants in rodents\cite{ref82,ref83,ref121}. The advantage is essentially zero electromagnetic footprint, the disadvantage is rapid 1/$d^2$ fall-off through pigmented and vascularized tissue.

\subsection*{Uplink modalities}

As opposed to downlink modalities, uplink communication focuses primarily on achievable data rates, within the available power budgets, to efficiently transmit clinical-rate neural data.
%from the implant to the external world.

\emph{Backscatter} uplinks re-radiate or re-modulate the energy of an external interrogator, eliminating the implant's local oscillator and power amplifier, typically the dominant DC power consumers in an active communication\cite{ref23,ref24,ref25,ref43,ref93,ref120}. Ultrasonic backscatter in neural dust pushes uplink energy below 1 nJ/b at $\leq$100 kbps\cite{ref23,ref24,ref25}. Inductive load-shift-keying (LSK) has been shown at 30 Mbps at <1 pJ/b for 64-channel $\mu$ECoG\cite{ref120,ref35,ref95}. The fundamental cost is external receiver's complexity: extracting a sub-percent modulation against the full carrier demands $\geq$80 dB self-interference cancellation in most cases, along with a much lower maximum data-rate possible (or channel capacity) due to incurring roundtrip loss.

\emph{Active RF narrowband} transmitters in MedRadio and the 2.4-GHz ISM band remain the default for clinically commercialized neuromodulators (see Section 3) because they leverage mature and compliant frameworks\cite{ref9,ref69,ref77}. BLE-compliant transmitters under 7 nJ/b 
%and 31.6-$\mu$W BLE-conformant designs
have been shown \cite{ref39, ref115}, making them viable for medium-rate sub-Mbps uplinks.
%Wake-up receivers at --80 dBm sensitivity and $<$250 nW\cite{ref116,ref117} power help reducing the  \emph{average} communication cost.

\emph{Impulse-radio ultra-wideband} (IR-UWB) telemetry avoids high-power clocking and power amplification by transmitting nanosecond-scale pulses whose spectrum spans 3.1--10.6 GHz\cite{ref43,ref44}. Prior duty-cycled all-digital pulse generators\cite{ref43,ref44} achieved 47 pJ/pulse, and recent transcutaneous IR-UWB has reached 1.66 Gbps at 5.8 pJ/b with hybrid OOK/PPM modulation\cite{ref41,ref42}. IR-UWB is now the highest-throughput uplink modality  \cite{ref35,ref40,ref41,ref42}, consistent with FCC Part 15 emission masks, although tissue attenuation above 5 GHz is $\sim$15 dB worse, as compared to air \cite{lei20252} at 10-mm depth.

\emph{EQS-BCC} uses E-field communication that leverages the conductive, dispersive volume of the brain-channel as a moderate-loss wideband channel from $<$1 MHz to $\sim$100 MHz\cite{ref45,ref46,ref47,ref48,ref49,ref50,ref52,ref54,ref57,ref144}. Because the dominant fields are quasi-static rather than radiating, channel loss at lower frequencies is governed by termination impedances and tissue conductivity rather than by SAR-limited propagation.
%and total path loss between an intracranial implant and a wearable on the arm can be $<$50 dB even across the skull\cite{ref144,ref54}. %BodyWire\cite{ref45} achieved 6.3 pJ/b at 30 Mbps between wearables; BP-QBC\cite{ref54,ref144} extends the same physics into the brain at 1.15 $\mu$W total implant power.
EQS-BCC also possesses an intrinsic \emph{signal-privacy} property: the field is confined to within a few centimeters of the body, limiting eavesdropping range to $\leq$15 cm\cite{ref50,ref142}. Similar to mid-field H-field techniques at higher frequencies than the quasistatic region, mid-field E-field BCC communication is also possible \cite{sarkar2025body}. %Galvanic operation removes the need for a return-path capacitor and is preferred for fully encapsulated implants\cite{ref48}.

\subsection*{Wireless Link budgets, Shannon channel capacity, and the Chu limit}

This section will explain the three fundamental considerations that govern wireless telemetry: link budgets, channel capacity and geometry constraints. As an example for link budgeting, we start with inductive coupling, which is relatively well understood for implantable wireless links. For a transcutaneous inductive link at depth \emph{d}, PTE scales as:

\begin{equation}
\mathrm{PTE} \approx \frac{k^2 Q_1 Q_2}{(1 + \sqrt{1 + k^2 Q_1 Q_2})^2},
\end{equation}

where \emph{Q}$_{1}$ and \emph{Q}$_{2}$ are coil quality factors\cite{ref19,ref73,ref74}. Because \emph{k} falls roughly as (\emph{r}$_{2}$/\emph{d})$^{3}$ for unmatched diameters, miniaturizing the implant coil to \emph{r}$_{2}$ < 2 mm at \emph{d} > 10 mm typically limits PTE below 1\% even with \emph{Q} > 100\cite{ref15,ref93,ref95}.
%For deep mm-sized implants, the optimum carrier frequency moves into the low-GHz range (~1--2 GHz)\cite{ref22}, where the wavelength in tissue approaches the implant scale, and evanescent coupling efficiency exceeds the quasi-static maximum by up to 30 dB\cite{ref20,ref21,ref22}.
The channel loss eventually determines the SNR for a particular transmitted power within safety limits, which in turn determines link budgets.

For ultrasound, the path loss is dominated by impedance mismatch at air--skin, skull, and tissue interfaces and by frequency-dependent absorption (~1 dB/cm/MHz in soft tissue), often favoring operation around or below 1 MHz \cite{ref23,ref24,ref25,ref26,ref27,ref105,ref106,ref107,ref108,ref109}. %The skull alone introduces 10--20 dB of additional loss and severe phase aberration above 2 MHz, motivating hybrid RF-to-ultrasound relays at the skin surface\cite{ref79}.
Magnetoelectric (ME) coupling, in contrast, traverses heterogeneous tissue with negligible loss because its acoustic transduction occurs only \emph{inside} the implant, and its kHz-range driving field encounters tissue conductivities of less than 1 S/m \cite{ref28,ref29,ref30,ref112}.
%The relative magnetic permeability of 1 for the tissue indicates that the low-frequency magneto-quasistatic fields propagating through the tissue for ME coupling experiences significantly low loss.
%Now, although understanding of such loss mechanisms, and the resulting SNR availability requires highly complex and multi-faceted theoretical derivation for every physical mechanism, the high-speed communication requirements of the system does not only depend on the SNR. Channel capacity in the weakly coupled regime obeys Shannon's classical bound $C = B\log_{2}(1+\mathrm{SNR})$, which values bandwidth far more over SNR.
However, while loss mechanisms determine the available SNR at the receiver, high‑rate neural uplinks are ultimately limited by Shannon capacity:

\begin{equation}
C = B\log_{2}(1+\mathrm{SNR}),
\label{eq:Shannon}
\end{equation}

Interestingly, Eq. \ref{eq:Shannon} rewards bandwidth linearly, while being only a logarithmic function of SNR. In the weakly coupled regime, where SNR is often modest but bandwidth can vary by orders of magnitude across modalities, capacity tends to be bandwidth‑limited rather than SNR‑limited.
Evidently, every term in the above equation is heavily constrained: bandwidth \emph{B} is limited by carrier choice and tissue dispersion; SNR is set by transmit power (itself bounded by SAR and temperature limits) and the noise floor at the receiver\cite{ref62,ref63,ref64,ref65,ref66}. %Empirical implant-to-surface path-loss models in the MedRadio band give \emph{PL}(d) = 47.14 + 4.26\textperiodcentered 10 log$_{10}$(d) dB with shadowing $\sigma$ $\approx$ 7.85 dB for deep tissue\cite{ref64} while measured EQS-HBC channels exhibit <30 dB total path loss between an intracranial implant and a wearable receiver below 30 MHz\cite{ref46,ref54,ref144}.
%Crucially, the \emph{useful} bandwidth in the tissue often shrinks faster than the available circuit bandwidth: dielectric dispersion broadens nanosecond IR-UWB pulses by 30--80\% after 10 mm of tissue\cite{ref41,ref42}, while inductive links with high \emph{Q} trade bandwidth for efficiency.
In Section 4, we will see how all five modalities appear bandwidth‑limited in the uplink. \emph{\ul{Resonant NIC/US/ME links sacrifice bandwidth to maximize PTE, whereas UWB and BCC accept higher path loss but leverage 100–700 MHz of effective bandwidth to maximize channel capacity.}}
A practical rule for the weakly coupled regime is that effective channel capacity per unit implant volume scales as the \emph{product} of channel coupling efficiency and the carrier bandwidth that survives tissue dispersion,   not simply as the hardware clock rate. The SNR and bandwidth scaling laws in Part I of Table \ref{tab:combined} are direct algebraic expressions of this rule for each modality.

A further subtlety is that the implant antenna or transducer is electrically small (\emph{kL} $\ll$ 1, where \emph{k} $= \frac{2\pi}{\lambda}$ is the wavenumber and \emph{L} the largest dimension) and therefore subject to the Chu limit \cite{chu1948physical} on minimum quality factor and hence, maximum bandwidth:

\begin{equation}
Q_\mathrm{min} \geq \frac{1}{(kL)^3} + \frac{1}{(kL)},
\label{eq:Chu}
\end{equation}

%For a 2-mm dipole at 915 MHz in muscle, the radiation efficiency rarely exceeds 1--3\% even after dielectric matching\cite{ref78}. Magnetic dipoles fare somewhat better because the lossy tissue dielectric does not couple to them as strongly\cite{ref70,ref71}. 
At 4 GHz frequency inside the brain ($\epsilon_r \approx$ 50, $\lambda_\mathrm{tissue} \approx$ 11 mm), a 1 mm implant has \emph{kL} $\approx$ 0.57 and $Q_\mathrm{min}$ $\approx$ 9.3. The resulting fractional bandwidth is $FBW_\mathrm{max} = 1/Q_\mathrm{min}$ $\approx$ 10.7\%, which is only 428 MHz at 4 GHz center frequency. This is less than 25\% of the target UWB bandwidth, meaning a 1 mm implant cannot maintain the wideband characteristic of UWB due to the Chu limit.
The miniaturization analysis presented in the supplementary section uses the Chu limit to show that for compact, flat‑die UWB antennas, the radiation bandwidth remains orders of magnitude larger than narrowband modalities, even when shrunk to 1 mm$^{3}$.
In Section 4, we consider five explicit physical constraints that determine channel capacity: (1) tissue propagation loss, (2) geometric and depth‑dependent coupling, (3) transducer electro‑mechanical efficiency, (4) impedance mismatch, and (5) safety‑limited transmit power. This decomposes `weak coupling' as an explicit, computable function of these parameters, enabling the cross‑modality comparison in Table \ref{tab:combined}.
%This paper will provide a five-term decomposition of weak coupling into tissue propagation loss, geometric spreading/coupling, transducer electro‑mechanical efficiency, impedance mismatch, and safety-induced power back‑off, turning ``weak coupling'' into an explicit, computable framework across modalities.
%We quantitatively separate tissue loss across modalities using Cole–Cole dispersion and skin-depth analysis, showing how inductive and ME links are geometry‑limited while ultrasound is limited by skull absorption. Depth and geometry-dependent channel capacity scaling laws (e.g., \emph{d}$^{-3}$ for inductive/ME vs. \emph{d}$^{-3/2}$ for ultrasound) are derived, making the weakly coupled regime a set of depth‑specific operating regimes rather than a single qualitative category.
%Pulling these constraints together, the achievable downlink PTE and uplink channel capacity of a weakly coupled implant are best understood as Pareto fronts in a five‑dimensional space: tissue loss, geometric coupling, transducer efficiency, impedance mismatch and safety‑limited transmit power. Channel capacity must therefore be computed end‑to‑end, including dispersion, geometry, transducer loss and receive‑side SNR under a consistent noise floor and termination. Section 4 implements this framework for NIC, US, ME, BCC and UWB, normalizing implant volume, depth and safety envelope so that the resulting Shannon capacities can be compared on a common energy‑per‑bit and data‑rate plane.
The safety considerations of the different modes are discussed in the Supplementary section.

% \begin{boxsection}
% \textbf{Box 1 | Weakly coupled link budget (worked example)}

% Consider a fully implanted intracortical recorder at 10-mm depth with a 2-mm-diameter receive coil, communicating to a 30-mm-diameter external coil. At 13.56 MHz with \emph{Q}$_{1}$ = 200 and \emph{Q}$_{2}$ = 50, \emph{k} $\approx$ 0.005 yields PTE $\approx$ 0.6\%; delivering 1 mW to the implant requires ~170 mW radiated and produces a peak SAR of ~0.9 W kg$^{-1}$ at the skull--cortex boundary, within IEEE C95.1-2019 limits but with only 4 dB headroom\cite{ref19,ref65,ref90}. Moving to mid-field operation at 1.6 GHz with a Poon-style multi-port external array raises link efficiency to ~2\% for the same implant size\cite{ref20,ref21,ref22} trading a higher SAR density at the skull surface for an order-of-magnitude smaller receive aperture and >100 Mb s$^{-1}$ achievable uplink at <10 mW implant DC power\cite{ref41,ref90}.
% \end{boxsection}

\subsection*{Analysis of state-of-the-art wireless link implementations}

Table \ref{tab:modified_final_version} demonstrates a comparative study of selected in-vivo wireless implant telemetry demonstrations, categorized into NIC, RF-NB, UWB, US, ME and BCC systems. Papers published during 2016-2026 are considered, and emphasis has been placed on literature that includes in-vivo and/or relevant in-vitro studies.

%=========================================================
% Column and helper definitions
%=========================================================

% Left-aligned fixed-width column.
\newcolumntype{P}[1]{>{\RaggedRight\arraybackslash\hspace{0pt}}p{#1}}

\providecommand{\nr}{-}

% Citation-only cell.
% Paper identifier text is removed; only citation remains.
\newcommand{\CIT}[1]{%
\parbox[t]{\linewidth}{%
\RaggedRight
\fontsize{8pt}{5.5pt}\selectfont
\cite{#1}%
}%
}

% Left-aligned wrapped normal text cell.
\newcommand{\Tcell}[1]{%
\parbox[t]{\linewidth}{%
\RaggedRight
#1%
}%
}

% Smaller left-aligned header cell.
\newcommand{\Hcell}[1]{%
\parbox[t]{\linewidth}{%
\RaggedRight
\fontsize{8pt}{5.50pt}\selectfont
\bfseries
#1%
}%
}

% Reusable smaller left-aligned header row.
\newcommand{\HeaderRow}{%
\centering
%\Hcell{Mod.} &
\Hcell{Ref.} &
\Hcell{Venue} &
\Hcell{Year} &
\Hcell{UL} &
\Hcell{DL} &
\Hcell{WPT} &
\Hcell{Channels} &
\Hcell{Carrier /\\frequency} &
\Hcell{DR\\Mb/s} &
\Hcell{Tx\\pJ/b} &
\Hcell{Rx\\pJ/b} &
\Hcell{Depth\\mm} &
\Hcell{Vol.\\mm$^3$} &
\Hcell{BER/PER}%
}

%=========================================================
% Portrait centered 15-column longtable
% Citation-only reference column, wider spacing
%=========================================================
%\clearpage
%\onecolumn
%\newgeometry{left=0.7in,right=1in,top=1in,bottom=1in}

{
{\fontsize{8pt}{5.5pt}\selectfont
\setlength{\tabcolsep}{1.35pt}
\renewcommand{\arraystretch}{1}
\setlength{\baselineskip}{0pt}

% Center the longtable on the page
% \setlength{\LTleft}{\fill}
% \setlength{\LTright}{\fill}
% Let longtable extend into left and right margins
\setlength{\LTleft}{-0.15in}
\setlength{\LTright}{-0.15in}

\sloppy
\emergencystretch=18em
\hyphenpenalty=50
\exhyphenpenalty=50

\begin{longtable}{@{}%
P{1cm}
P{1.15cm}
P{0.64cm}
P{0.8cm}
P{0.8cm}
P{0.8cm}
P{2cm}
P{2.2 cm}
P{0.9cm}
P{0.75cm}
P{0.75cm}
P{1.2cm}
P{0.9cm}
P{2cm}
%P{2cm}
@{}}

\caption{\textbf{Selected in-vivo wireless implant telemetry demonstrations (by modality)}}
\label{tab:modified_final_version}\\

\toprule
\HeaderRow \\
\midrule
\endfirsthead

\caption[]{\textbf{Selected in-vivo wireless implant telemetry demonstrations} (continued).}\\

\toprule
\HeaderRow \\
\midrule
\endhead

\midrule
\multicolumn{14}{r}{\emph{Continued on next page}}\\
\endfoot

\bottomrule
\endlastfoot

%========================================================================================
% Inductive
%========================================================================================
\multicolumn{14}{@{}l@{}}{\textit{Inductive links (Ind./NIC)}} \\
\midrule

%Ind. &
\cite{park2025enhanced} &
ISSCC &
2025 &
- &
Ind. &
Ind. &
- &
6.8/7.2 MHz&
1 &
\nr &
\nr &
\Tcell{5\\(porcine head)} &
\nr & BER $<$ 1e-7 \\

%Ind. &
\cite{park2021wireless} &
TBCAS &
2021 &
- &
Ind. &
Ind. &
- &
6.5/7.5 MHz&
2.5 &
\nr &
\nr &
\Tcell{5\\(pork tissue)} &
\nr &
BER $\sim$4e-7 \\

%Ind. &
\cite{ref33} &
\Tcell{Nature\\Elec.} &
2021 &
Ind. &
Ind. &
Ind. &
\Tcell{multi-site/ networked} &
\Tcell{$\sim$ 1 GHz link frequency} &
10 &
\nr &
\nr &
\Tcell{5\\(rodent)} &
0.1 &
BER $\sim$4e-5 \\

%Ind. &
\cite{jia2019mm} &
TBCAS &
2019 &
Ind. &
Ind. &
Ind. &
\nr &
60 MHz &
0.05 &
\nr &
\nr &
\Tcell{7\\(sheep model)} &
9.3 &
- \\

%Ind. &
\cite{li2018200,8310298} &
\Tcell{TBCAS/\\ISSCC} &
2019 &
Ind. &
- &
- &
- &
\Tcell{200-MHz UL TX clock} &
200 &
1.5 &
\nr &
\Tcell{11.8\\(piglet skin and skull)} &
\nr &
\Tcell{BER $\sim$ 5e-11 through skull; $<$1e-12 in air gap} \\

%Ind. &
\cite{yu2022wireless} &
RFIC &
2022 &
Ind. &
ME &
ME &
\Tcell{multi-site/ networked} &
\Tcell{DL 0.340 MHz; UL 31 MHz} &
\Tcell{DL 0.0623;\\UL 0.040} &
\nr &
\nr &
\Tcell{20\\(pork tissue)} &
8.8 &
5.5e-5 UL \\

%Ind. &
\cite{jiang2016integrated} &
TBCAS &
2017 &
Ind. &
Ind. &
Ind. &
\Tcell{-} &
13.56 &
1.35 &
\nr &
\nr &
\Tcell{4\\(pork)} &
\nr &
\Tcell{BER $<$ 6e-8} \\

%Ind. &
\cite{yang2022neural} &
TBCAS &
2023 &
RF/ Ind. &
- &
RF/ Ind. &
1 &
6 MHz&
0.25 &
\nr &
\nr &
\Tcell{16\\(rodent)} &
0.4 &
\Tcell{BER = 8e-6 @16 mm} \\

%========================================================================================
% RF-NB
%========================================================================================
\midrule
\multicolumn{14}{@{}l@{}}{\textit{Narrowband RF (RF-NB) and backscatter-assisted RF}} \\
\midrule

%RF-NB &
\cite{huang202515} &
ISSCC &
2025 &
RF-NB &
Ind. &
Ind. &
\nr &
\Tcell{510 MHz WPT; 390 MHz UL backscatter} &
200 &
0.67 &
21.7 &
6.6 &
\nr &
\Tcell{UL BER $<$7.5e-5; DL BER $<$1e-5} \\

%RF-NB &
\cite{kampianakis2017dual} &
JRFID &
2017 &
RF-NB &
- &
RF-NB &
\Tcell{10 neural + 4 EMG channels} &
\Tcell{13.56 MHz WPT; 915 MHz backscatter} &
5 &
\nr &
\nr &
\Tcell{10\\(saline)} &
1313 &
\Tcell{0\% PER at up to 2.5 cm} \\

%========================================================================================
% UWB
%========================================================================================
\midrule
\multicolumn{14}{@{}l@{}}{\textit{Ultra-wideband RF (UWB)}} \\
\midrule

%UWB &
\cite{song20252} &
TCAS-II &
2025 &
UWB &
- &
- &
\nr &
6500--8500 MHz &
2540 &
1.22 &
\nr &
\Tcell{3\\(beef skin)} &
\nr &
\Tcell{BER $\sim$ 1e-3 @3.4 cm} \\

%UWB &
\cite{lei20252} &
JSSC &
2026 &
UWB &
- &
- &
\nr &
3100--7000 MHz&
2016 &
1.9 &
8.5 &
\Tcell{10\\(7-layer brain phantom)} &
\nr &
\Tcell{BER$\sim$ 1e-3 @40 cm at 2.016 Gbps and @120 cm at 1.344 Gbps} \\

%UWB &
\cite{ding202549} &
JSSC &
2025 &
UWB &
- &
- &
\nr &
4000--6000 MHz&
800 &
16.5 &
\nr &
\Tcell{15\\(chicken)} &
\nr &
\Tcell{BER$\sim$ 7e-4 @80 cm and 1e-3 @125 cm} \\

%UWB &
\cite{ding20253} &
JSSC &
2025 &
UWB &
- &
- &
\nr &
3000--6250 MHz&
1200 &
6 &
\nr &
\Tcell{15\\(meat)} &
\nr &
\Tcell{BER$<$4.17e-6 at 1.2 Gb/s} \\

%UWB &
\cite{ref41,9921401} &
\Tcell{ISSCC/\\JSSC} &
2022 &
UWB &
- &
- &
\nr &
6000--9000 MHz&
1660 &
5.8 &
\nr &
\Tcell{15\\(tissue)} &
\nr &
\Tcell{BER$\sim$ 2e-4} \\

%UWB &
\cite{lei20231} &
ISSCC &
2023 &
UWB &
- &
- &
\nr &
3100--5000 MHz&
1800 &
2.3 &
\nr &
\Tcell{18\\(pork)} &
\nr &
\Tcell{BER$\sim$ 1e-4 @15 cm and 1e-3 @20 cm} \\

%UWB &
\cite{jung2025wireless} &
\Tcell{Nature Elec.} &
2025 &
UWB &
UWB &
Ind. &
\Tcell{65,536; 1,024 simultaneous} &
\Tcell{4 GHz UWB; 13.56 MHz WPT; 700 MHz UWB BW} &
108.48 &
39 &
200 &
\Tcell{intra-\\cortical} &
7.2 &
- \\

%UWB &
\cite{liu2026compressive} &
TMTT &
2026 &
UWB &
- &
- &
\Tcell{384; Neuropixels-based telemetry} &
\Tcell{6--8.5 GHz IR-UWB} &
138 &
7.2 &
- &
\Tcell{15 mm\\(tissue)} &
\nr &
\Tcell{BER $\sim$1e-3 @660 cm air} \\

%========================================================================================
% Ultrasound
%========================================================================================
\midrule
\multicolumn{14}{@{}l@{}}{\textit{Ultrasonic links (US)}} \\
\midrule

%US &
\cite{ref23} &
Neuron &
2016 &
US &
- &
US &
\nr &
$\sim$1.85 MHz &
0.5 &
\nr &
\nr &
\Tcell{8.8\\(tissue)} &
2.4 &
- \\

%US &
\cite{hosur2023magsonic} &
TBioCAS &
2024 &
US/ ME &
- &
US/ ME &
\nr &
1.93 MHz&
0.1 &
190 &
\nr &
\Tcell{40\\(aqueous)} &
16.64 &
\Tcell{BER $\leq$1e-5 @40 mm, 100 kbps} \\

%US &
\cite{ref26} &
JSSC &
2019 &
US &
- &
US &
\nr &
1.78 MHz &
0.035 &
\nr &
1077 &
\Tcell{50\\(oil)} &
0.8 &
- \\

%US &
\cite{gao2025simultaneous} &
ASSCC &
2025 &
US &
US &
US &
\nr &
1 MHz&
0.3 &
6.3 &
\nr &
\Tcell{50\\(phantom)} &
\nr &
BER $\sim$1e-6 \\

%US &
\cite{chang201727} &
ISSCC &
2017 &
US &
US &
US &
1 &
\Tcell{US WPT $\sim$1 MHz; US UL $\sim$2.5 MHz} &
0.1 &
\nr &
\nr &
\Tcell{60\\(tissue)} &
30.5 &
\Tcell{BER $<$1e-4 @8.5 cm} \\

%========================================================================================
% ME
%========================================================================================
\midrule
\multicolumn{14}{@{}l@{}}{\textit{Magnetoelectric links (ME)}} \\
\midrule

%ME &
\cite{wang2025dual} &
\Tcell{ACM\\MobiCom} &
2025 &
ME &
- &
ME &
1 &
0.8 MHz UL &
0.06 &
\nr &
\nr &
\Tcell{10\\(pork tissue)} &
91 &
\Tcell{OOK BER 2e-4 @20 mm; BPSK EVM 20\% @50 mm} \\

%ME &
\cite{yu2021magnetoelectric} &
JSSC &
2022 &
- &
ME &
ME &
\Tcell{1, multi-site/ networked} &
0.33 MHz &
0.00516 &
\nr &
\nr &
\Tcell{20\\(pork tissue)} &
6.2 &
\nr \\

%ME &
\cite{ref111,ref28} &
\Tcell{ISSCC/\\TBCAS} &
2020 &
- &
ME &
ME &
1 &
0.25 MHz &
0.0078 &
\nr &
\nr &
\Tcell{30\\(hydra)} &
8.2 &
\nr \\

%ME &
\cite{chen2022wireless} &
\Tcell{Nature\\Biomed.\\Eng.} &
2022 &
- &
ME &
ME &
1 &
0.345 MHz &
0.0046 &
\nr &
\nr &
\Tcell{40\\(pork tissue)} &
95 &
- \\

%ME &
\cite{yu2022magnetoelectric} &
\Tcell{ACM\\MobiCom} &
2022 &
ME &
ME &
ME &
1 &
0.335 MHz &
0.06 &
\nr &
\nr &
\Tcell{15\\(pork tissue)} &
8.2 &
\Tcell{BER $\sim$ 1e-5 @15 mm} \\

%ME &
\cite{yu2024miniature} &
TBCAS &
2024 &
ME &
- &
ME &
\Tcell{LFP recording demo} &
\Tcell{ME low-frequency} &
0.01773 &
0.9 &
\nr &
\Tcell{20\\(pork tissue)} &
6.7 &
\Tcell{BER $\sim$ 8.5e-5 @50 mm} \\

%ME &
\cite{yu202433} &
ISSCC &
2024 &
ME &
ME &
ME &
1 &
0.331 MHz&
0.0177 &
0.9 &
\nr &
\Tcell{50\\(in PBS)} &
6.7 &
\Tcell{BER $\sim$ 8.5e-5 @50 mm} \\

%========================================================================================
% BCC
%========================================================================================
\midrule
\multicolumn{14}{@{}l@{}}{\textit{Body-channel communication/Brain-channel communication (BCC)}} \\
\midrule

%BCC &
\cite{lee2022miniaturized} &
JSSC &
2022 &
BCC &
- &
BCC &
4 &
\Tcell{Carrierless, 40.96 MHz clock} &
20.48 &
32 &
\nr &
\Tcell{5\\(in-vivo)} &
\nr &
BER $\sim$ 0.001 \\

%BCC &
\cite{shi2024spatially} &
TBCAS &
2024 &
BCC &
- &
- &
\Tcell{2 Tx couplers, 3 Rx couplers} &
\Tcell{Baseband; max. BW 360 MHz} &
270 &
3.4 &
3.7 &
7 &
\nr &
BER $<$ 1e-6 \\

%BCC &
\cite{shi2022galvanic} &
T-MTT &
2022 &
BCC &
- &
- &
\Tcell{MEA data link; no neural-AFE} &
\Tcell{Baseband impulse; 250 MHz channel characterization} &
250 &
2 &
\nr &
\Tcell{10\\(muscle)} &
\nr &
\Tcell{BER test $\sim$ 1e-6 (250 Mb/s)} \\

%BCC &
\cite{ref54} &
ESSCIRC &
2022 &
BCC &
BCC &
BCC &
16 &
\Tcell{$\sim$ATC 800 kSps, 10-12b/sample} &
9.6 &
1.1 &
33 &
50 &
5.5 &
\Tcell{DL BER 10$^{-3}$ (1 kbps)} \\

%BCC &
\Tcell{\cite{ref144,9492445}} &
\Tcell{Nature Elec./ VLSI} &
\Tcell{2023/\\2021} &
BCC &
BCC &
BCC &
\Tcell{1, Networked neural sensor/stimulator SoC} &
\Tcell{10 MHz EQS; $\sim$100 MHz flatband channel} &
10 &
52 &
16 &
\Tcell{60\\(in PBS)} &
5.54 &
\Tcell{DL BER $<$10$^{-3}$; UL loss $\sim$60 dB @55 mm} \\

%BCC &
\cite{chang2025rpg} &
CICC &
2025 &
BCC &
- &
- &
\Tcell{-} &
\Tcell{40--100 MHz carrier} &
25 &
\nr &
\nr &
10 &
\Tcell{72*} &
BER $\sim$ 6.20e-5 \\

%BCC &
\cite{shen2024battery} &
JSSC &
2024 &
BCC &
- &
BCC &
16 &
\Tcell{1 MHz WPT; 2 MHz passive-BCC UL} &
0.25 &
\nr &
\nr &
55 &
5.9 &
BER $\sim$ 4.4e-6 \\

%BCC &
\cite{11509547} &
CICC &
2026 &
\Tcell{BCC} &
- &
- &
\Tcell{1} &
\Tcell{40 MHz} &
80 &
\nr &
0.67 &
\Tcell{100 (in tissue)} &
\nr &
\Tcell{BER $=10^{-3}$ @ 80 Mb/s} \\
\bottomrule

\end{longtable}
}
}
\vspace{-7mm}
\noindent \scriptsize{*estimated from the information presented in the paper}
\vspace{4mm}
\normalsize

%\restoregeometry

\baselineskip=14pt

\begin{figure}[t]
   \centering
    \includegraphics[width=1\textwidth]{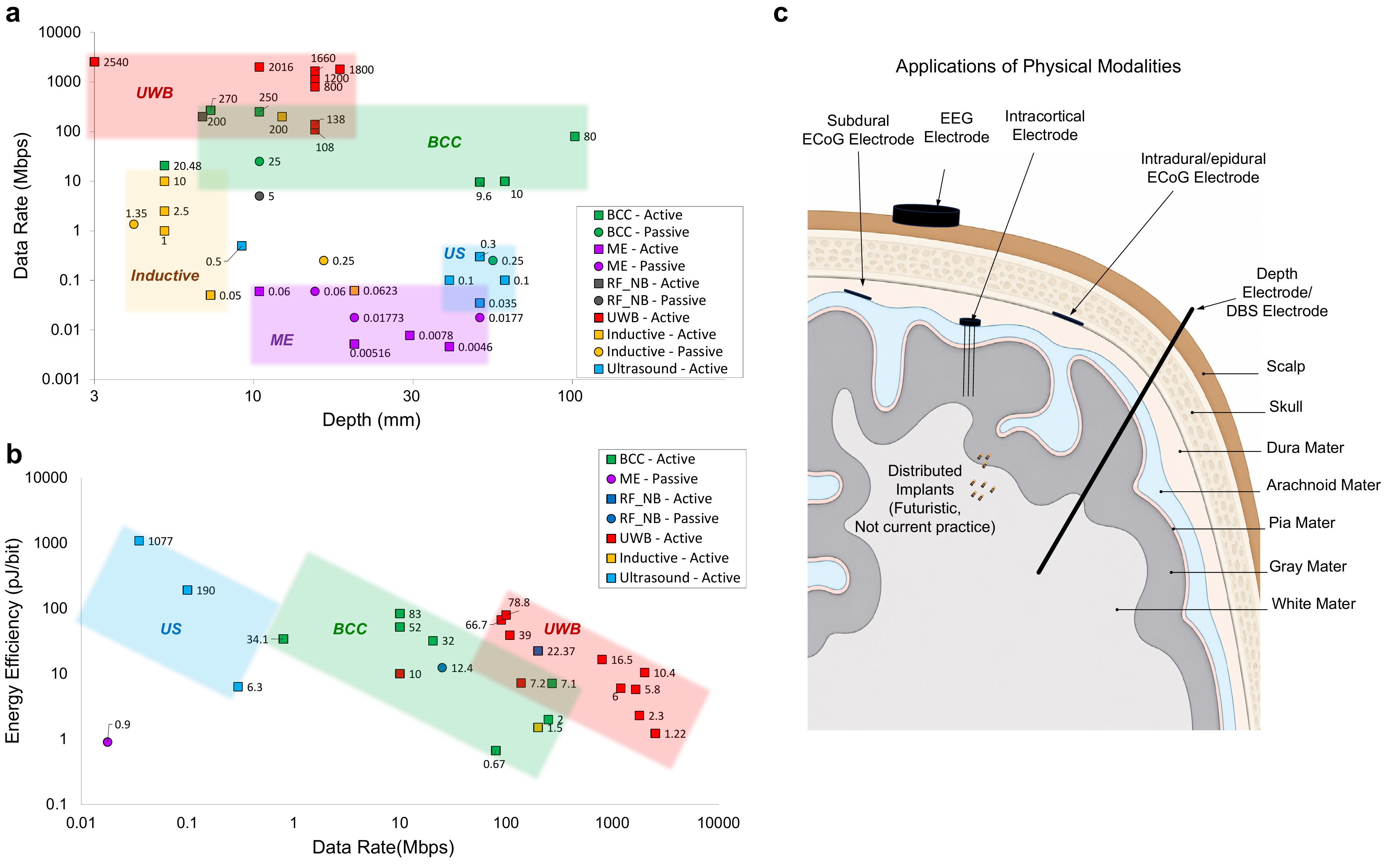}
    \caption{\textbf{Insights obtained from state-of-the-art wireless link implementations for Brain Implants. a,} Data Rate vs. Depth shows that wideband techniques (such as UWB or BCC) achieves significantly higher data rates than narrowband techniques. However, narrowband US or ME techniques, due to their low loss, can support higher depth (3-5 cm) for future applications. BCC has also been explored for $>$5 cm depth, and has proven to achieve higher data rates than conventional narrowband. The numerical values near the data points indicate the data rate in Mbps. \textbf{b,} Energy-efficiency vs. Data Rate, which demonstrates the improved energy efficiencies of wideband techniques (UWB or BCC) over narrowband. Interestingly, some of the narrowband techniques (such as US or ME) hits a limit in terms of geometries, and prove to be insufficient to support higher carrier frequencies or data rates. The numerical values near the data points indicate the energy-efficiency in pJ/b. \textbf{c,} Depending on the application and the surgical needs, one modality can be preferred over others, which dictates the commercial adoption of the technology.}
    \label{fig:SoA}
    \vspace*{-0in}
\end{figure}

Fig. \ref{fig:SoA}a-b exhibits the important insights obtained from Table \ref{tab:modified_final_version}, in terms of data rate vs. depth of the implant, and energy-efficiency vs. data rate of the implant. The plot of data rate versus implant depth shows that wideband techniques, such as UWB and BCC, achieve substantially higher data rates than narrowband techniques, with similar depth. However, narrowband ultrasound (US) and magnetoelectric (ME) approaches, owing to their low propagation loss, can support greater implant depths, typically around 3--5 cm, making them promising for future applications. BCC has also been investigated for depths exceeding 3--5 cm and has demonstrated higher data rates than conventional narrowband approaches. The energy-efficiency versus data rate plot, again, illustrates the superior energy efficiency of wideband techniques, including UWB and BCC, compared with narrowband approaches. Notably, some narrowband techniques, such as US and ME, encounter geometry-related limitations and may be insufficient for supporting higher carrier frequencies or higher data rates, as will be seen in Section 4. Depending on the application and surgical envelope, one modality can be preferred over others. Fig. \ref{fig:SoA}c shows the different commercially available device types, which is described in the commercial landscape of Section 3.
%========================================================================================
\section*{3. Commercial Landscape}

From the patient/user's standpoint, the invasive bidirectional BCI landscape is best read along one axis: the surgical envelope, which fixes the available volume, the thermal headroom, and the link-energy regime that the architecture must respect. Today's wireless implant links operate at $\sim$100\,pJ/b to 10\,nJ/b across BLE, UWB, mid-field, inductive, ultrasonic, and magnetoelectric demonstrations\cite{ref10,ref20,ref21,ref22,ref28,ref40,ref41,ref42}, which forces aggressive on-implant compression across nearly all commercial architectures; the sub-10\,pJ/b \emph{need} and sub-1\,pJ/b \emph{aspiration} are the thresholds at which the computation may move off-implant for $\sim$100 Mbps and $\sim$1 Gbps links, repectively. We organize the current landscape by surgical envelope, citing the link-layer nuance for each platform in Table~\ref{tab:commercial}.
%and exclude peripheral neuromodulators and retinal prostheses as reviewed elsewhere\cite{ref88,ref98}.

%\input{tables/commercial-table}
% 
%========================================================================================
%========================================================================================
%========================================================================================

%---
% Commercial Landscape table, Nature-format, link-energy-centric
% v4: tightened cell text; structure, citations, columns, and font settings unchanged

{
{\fontsize{8pt}{5.5pt}\selectfont
\setlength{\tabcolsep}{2.2pt}
\renewcommand{\arraystretch}{0.8}
\setlength{\baselineskip}{0pt}
\setlength{\LTleft}{0pt}
\setlength{\LTright}{0pt}

\begin{longtable}{@{}P{2.1cm}P{1.9cm}P{1.7cm}P{1.5cm}P{2.0cm}P{2.2cm}P{1.8cm}P{1.5cm}@{}}
%\centering
\caption{\textbf{Commercial and clinical-stage invasive bidirectional brain interfaces, organized by surgical envelope.} Entries grouped by recording or stimulation site, ordered by clinical maturity. \emph{Depth / location} fixes the surgical envelope; \emph{Data-rate need} is the aggregate uplink the application demands (raw or post-compression as deployed); \emph{Link-energy regime} positions each platform against the manuscript's axis: today $\sim$100\,pJ/b to 10\,nJ/b, need sub-10\,pJ/b, aspiration sub-1\,pJ/b.
%Peripheral neuromodulators and retinal prostheses are reviewed elsewhere.
Abbreviations: BLE: Bluetooth Low Energy; UWB: ultra-wideband; ME: magnetoelectric; NIR: near-infrared; FIH: first-in-human; IDE: investigational device exemption; BDD: breakthrough device designation; LFP: local field potential; HNS: hypoglossal nerve stimulator; SCS: spinal cord stimulator.}
\label{tab:commercial}\\
% \scriptsize
% \setlength{\tabcolsep}{3pt}
% \renewcommand{\arraystretch}{1.15}
% \begin{tabular}{@{}p{2.2cm}p{1.6cm}p{1.8cm}p{1.6cm}p{1.9cm}p{2.2cm}p{1.4cm}p{1.4cm}@{}}
\toprule
\textbf{Platform} & \textbf{Depth / location} & \textbf{Application} & \textbf{Channels} & \textbf{Data-rate need} & \textbf{Link layer} & \textbf{Link-energy regime} & \textbf{Status (year)} \\ \addlinespace[5pt]
\midrule
\multicolumn{8}{@{}l@{}}{\textit{Intracortical, high-channel-count penetrating}} \\
\midrule
Neuralink N1\cite{ref10,ref41} & Penetrating, motor cortex & Motor / speech BMI & 1{,}024 electrodes/ 64 threads & few-hundred kbps (compressed); raw $\sim$2 to 25\,Mbps & 2.4\,GHz BLE; inductive recharge & $\sim$100\,pJ/b to nJ/b & FIH 2024; IDE \\ \addlinespace[5pt]
Paradromics Connexus\cite{ref12} & Penetrating + chest module & Motor / speech BMI & 421 clinical microelectrodes, earlier $\geq$1{,}600 microwires & gigabit-class raw & Tethered today; RF planned & Wired today & FIH 2025; IDE 2025 \\ \addlinespace[5pt]
Blackrock NeuroPort / Utah\cite{ref11,ref15} & Penetrating, motor cortex & Motor BMI; research & 96 to 192 & few Mbps raw; kbps wireless variant & Mostly tethered; wireless home-use\cite{ref11} & Wired today & Long-running clinical \\ \addlinespace[5pt]
Kampto Neurotech BISC\cite{ref88} & Penetrating, cortex & High-density recording & 65{,}536 sites & 100\,Mbps (near-raw) & Custom high-rate wireless & Sub-10\,pJ/b implied & Announced 2025 \\ \addlinespace[5pt]
CorTec Brain Interchange\cite{ref122} & Penetrating + chest IPG & Closed-loop record + stim & 32 channels & sub-Mbps (event-driven) & Inductive recharge + RF to chest IPG & nJ/b class & FIH 2025; BDD 2026 \\ \addlinespace[5pt]
\midrule
\multicolumn{8}{@{}l@{}}{\textit{Cortical-surface ECoG and $\mu$ECoG}} \\
\midrule
Precision Layer 7-T\cite{ref13,ref89} & Subdural + external skull receiver & Motor / speech BMI; mapping & 1{,}024 to 4{,}096 (multiple array) & tens of Mbps raw & External receiver (off-implant) & Off-implant by design & 510(k) 2025 \\ \addlinespace[5pt]
INBRAIN graphene\cite{ref89} & Epicortical / subdural & Adaptive DBS; mapping & Tens of (graphene $\mu$ECoG) & sub-Mbps to few Mbps & RF telemetry to external puck & nJ/b class & FIH 2024; enrol. complete 2026 \\ \addlinespace[5pt]
NeuroXess flexible $\mu$ECoG & Subdural, cortical surface & Motor BMI; mapping & High-density array & tens of Mbps projected & RF telemetry & nJ/b class & FIH-stage \\ \addlinespace[5pt]
NeuroOne Evo\cite{ref13} & Subdural and sEEG & Mapping; ablation & 32 to 64 / array & kbps to Mbps & Wired clinical & Wired & FDA-cleared \\ \addlinespace[5pt]
Ad-Tech, DIXI, PMT, Integra, Stryker & Subdural / depth & Pre-surgical mapping & 32 to 256 / array & kbps to Mbps & Wired clinical & Wired & Established \\ \addlinespace[5pt]
\midrule
\multicolumn{8}{@{}l@{}}{\textit{Endovascular and minimally invasive}} \\
\midrule
Synchron Switch\cite{ref8,ref9} & Intravascular + chest transmitter & Discrete-command BMI & 16 stentrode & tens of kbps & 2.4\,GHz ISM-band, chest to wearable & nJ/b; bandwidth-light & FIH; COMMAND met 2024 \\ \addlinespace[5pt]
\midrule
\multicolumn{8}{@{}l@{}}{\textit{Sub-scalp EEG monitors}} \\
\midrule
Epiminder Minder & Sub-scalp, behind-ear & Long-term EEG (epilepsy) & 6 to 8 & kbps; event-driven & Inductive-class, coil to behind-ear & nJ/b (conservative) & FDA De Novo 2025 \\ \addlinespace[5pt]
UNEEG SubQ / EpiSight & Sub-scalp & Long-term EEG & 1 to 3 & kbps; episodic upload & Inductive subcutaneous & nJ/b class & CE-marked \\ \addlinespace[5pt]
\midrule
\multicolumn{8}{@{}l@{}}{\textit{Brain-spine and hybrid bidirectional loops}} \\
\midrule
Onward ARC-BCI\cite{ref136,ref137} & Cortical recorder + lumbar SCS & Locomotor restoration & WImagine + ARC-IM & low payload; latency-bound & Inter-implant coordination link & nJ/b; latency-binding & FIH 2023; trials \\ \addlinespace[5pt]
CorTec Brain Interchange\cite{ref122} & Penetrating + chest IPG & Closed-loop research & 32 to 64 & sub-Mbps (event-driven) & Inductive recharge + RF & nJ/b class & FIH 2025; BDD 2026 \\ \addlinespace[5pt]
\midrule
\multicolumn{8}{@{}l@{}}{\textit{Free-floating and mm-scale}} \\
\midrule
Iota neural dust\cite{ref23,ref27,ref105} & Peripheral nerve (sub-mm$^3$) & PNS record / stim & Single site & $<$10\,kbps (backscatter) & Ultrasonic + backscatter & $<$1\,nJ/b; rate-bound & Preclinical \\ \addlinespace[5pt]
Motif DOT (ME)\cite{ref28,ref30,ref111} & Sub-dural / epidural & Targeted stimulation & Single site & Minimal uplink & ME power; backscatter uplink & Backscatter regime & FDA IDE 2026 \\ \addlinespace[5pt]
ni2o KIWI & Minimally invasive mm-scale & Stim / record & Undisclosed & Power-budget-limited & Wireless (undisclosed) & Backscatter / harvested & Preclinical \\ \addlinespace[5pt]
Subsense MENP & ME nanoparticles (non-implanted) & Trans-cranial stim & - & - & ME coupling & Backscatter regime & Research \\ \addlinespace[5pt]
Berkeley/UCSF neural dust\cite{ref23,ref24,ref25,ref26} & Sub-mm$^3$ motes (PNS / CNS) & Record, stim, multimodal & Single/mote & $<$kbps to kbps & Ultrasonic backscatter & $<$1\,nJ/b & Preclinical \\ \addlinespace[5pt]
NIR-optical sub-dural\cite{ref82,ref83,ref102,ref104,ref121} & Sub-dural photo-implant & Record; bidir. stim & Tens & set by 0.1 to 1\,mW harvested & VCSEL-to-photodiode optical & Off-implant; rate-bound & Preclinical \\ \addlinespace[5pt]
\midrule
\multicolumn{8}{@{}l@{}}{\textit{Established clinical neuromodulators (link-conservatism baseline)}} \\
\midrule
Medtronic Percept RC/PC + BrainSense\cite{ref84,ref146} & Deep brain + chest IPG & Adaptive DBS (PD, ET, OCD) & 4/lead $\times$ 2; LFP & sub-Mbps (LFP logs) & Inductive + sub-Mb MedRadio\cite{ref69} & nJ/b (conservative) & FDA-approved (RC 2024) \\ \addlinespace[5pt]
Boston Sci Vercise Genus\cite{ref84} & Deep brain + chest IPG & Directional DBS (PD, ET) & 8 directional/lead & sub-Mbps (telemetry) & Inductive + MedRadio & nJ/b class & FDA-approved \\ \addlinespace[5pt]
Abbott Infinity / Liberta RC\cite{ref84} & Deep brain + chest IPG & Directional DBS & 8 directional/lead & sub-Mbps (telemetry) & Inductive + MedRadio & nJ/b class & FDA-approved \\ \addlinespace[5pt]
Aleva directSTIM & Deep brain + chest IPG & Directional STN DBS & Multi-segment leads & sub-Mbps (telemetry) & Inductive + MedRadio & nJ/b class & CE-marked \\ \addlinespace[5pt]
NeuroPace RNS\cite{ref122} & Cortical/depth + cranial IPG & Closed-loop epilepsy & 4/lead $\times$ 2; LFP & kbps event logs & Magnetic-induction wand & nJ/b (per-session) & FDA-approved \\ \addlinespace[5pt]
Neuracle NEO & Extradural cortical + IPG & Closed-loop motor BMI & Extradural electrodes & sub-Mbps (compressed) & Inductive + RF telemetry & nJ/b class & NMPA-approved invasive BCI (2026) \\ \addlinespace[5pt]
Cochlear / Adv. Bionics / MED-EL\cite{ref71,ref72,ref73,ref74} & Cochlea + behind-ear & Auditory restoration & 12 to 22 stim & audio-rate downlink (mW) & Resonant-inductive 1 to 13.56\,MHz & $\mu$J/b powered downlink & $>$10$^{5}$ shipped \\ \addlinespace[5pt]
Inspire HNS\cite{ref84} & Hypoglossal + chest IPG & OSA neuromodulation & Stim lead & sub-Mbps (telemetry) & Inductive + MedRadio & nJ/b class & FDA-approved \\
\bottomrule
\bottomrule
%\end{tabular}
\end{longtable}
}

\baselineskip=14pt

\subsection*{Intracortical, high-channel-count invasive platforms}

Penetrating platforms share an on-board ASIC plus wireless-link, but differ in the link modality, which dictates their compression required. Neuralink N1 ($\sim$1024 polyimide threads, transcutaneous, inductive recharge\cite{ref10,ref41}, 2.4-GHz BLE) sits in the $>$ 1 nJ/b regime that forces aggressive on-implant compression\cite{ref123,ref124,ref125,ref126}. Paradromics Connexus ($\geq$1600-microwire pedestal, sub-clavicular transceiver, tethered\cite{ref12}) takes a clean \emph{wired first, wireless later} approach. Blackrock NeuroPort\cite{ref11,ref15} remains the longest-running Utah-array platform and demonstrated the first wireless intracortical home use. Kampto BISC is the link-energy outlier: 100\,Mbps from 65536 sites implies sub-100\,pJ/b\cite{ref88}, the regime in which raw streaming becomes feasible at the same surgical envelope. CorTec Brain Interchange is a bidirectional record-and-stimulate system on inductive recharge, plus RF telemetry to a chest IPG-class event-driven transceiver\cite{ref122}. Across such devices, the wireless layer spans BLE-class narrowband to custom IR-UWB at $\sim$100\,Mbps\cite{ref40,ref41,ref42}, and the spread in link energy is the best predictor of how much feature commitment is baked in. These devices penetrate 1.5 to 4\,mm into motor or speech cortex, and communicate with skull-mounted ASIC and antenna (Neuralink), or with cranial recorder/chest IPG-class module (Paradromics/CorTec). Raw data-rates range within 2--100\,Mbps for 256--1024 channels, projected 1--10\,Gbps for million-channel platforms\cite{ref89,ref141}; today's deployed uplinks deliver few-hundred kbps to $\sim$100\,Mbps depending on on-implant compression achieved.

\subsection*{Cortical-surface ECoG and \texorpdfstring{$\mu$}{u}ECoG}

Cortical-surface platforms relax the surgical envelope but inherit the link-energy problem as penetrating electrodes. Precision Neuroscience’s Layer 7-T interface moves link energy off the implant by routing wired signals to an external skull-mounted receiver\cite{ref13,ref89}; a 5376-channel head-mounted $\mu$ECoG demonstration recently achieved $>$100\,Mbps in the same family\cite{ref89}. INBRAIN's graphene cortical interface and NeuroXess's flexible $\mu$ECoG both utilize conventional nJ/b-class RF telemetry today because channel count is modest, but are positioned to scale into higher channel-counts. NeuroOne Evo\cite{ref13} and the established wired sEEG/subdural-grid suppliers (Ad-Tech, DIXI, PMT, Integra, Stryker) define the rate, latency, and reliability budget against which wireless successors are compared. These devices are epicortical and subdural, with no parenchymal penetration. A skull-mounted external receiver (Precision Layer 7-T) or cranially-routed  behind-ear or chest-pocket transceiver (INBRAIN, NeuroXess) receives the data at tens of Mbps for current 1024-contact arrays and $>$100\,Mbps for 5000-channel head-mounted demonstrations\cite{ref89}.

\subsection*{Endovascular and minimally invasive}

The Synchron Switch\cite{ref8,ref9} places a stentrode in the superior sagittal sinus, and communicates through a chest-implanted transmitter to a wearable receiver via 2.4\,GHz ISM-band. The clinical advantage (no craniotomy) trades-off data rate: intravascular ECoG-class recording at tens of kbps is two to three orders below intracortical needs.
%so the chest pocket grants two to three orders more energy budget and the link-energy frontier is non-binding today.
%The \emph{split} architecture recapitulates the cochlear/DBS pattern\cite{ref71,ref72,ref73,ref74,ref84} and is the most repeated motif below. \emph{Summary:} Depth: intravascular (superior sagittal sinus), no craniotomy; chest sub-clavicular transmitter. Location: split between intracranial vessel and chest IPG pocket. Data-rate need: tens of kbps aggregate, dominated by event-driven discrete commands rather than continuous high-bandwidth neural traffic; this is two to three orders of magnitude below the intracortical and cortical-surface categories and is why the link layer is non-binding today.

\subsection*{Sub-scalp EEG monitors}

Sub-scalp EEG implants form a clinical family distinct from cortical BCIs. Targeted toward long-term ambulatory monitoring (one to ten channels, kbps aggregate data rate), Epiminder Minder (subcutaneous behind-ear lead, sub-scalp coil) and UNEEG SubQ/EpiSight (one to three channels) use a conservative inductive-link with event-driven telemetry and on-board buffering. Architecturally they show that
%the \emph{split-topology, conservative-link} pattern is not neuromodulator-specific:
when recording bandwidth is bounded by the surgical envelope, on-implant compression is replaced by on-implant storage plus episodic telemetry.
%\emph{Summary:} Depth: sub-scalp, above the skull. Location: behind-ear (Epiminder) or subcutaneous cranial (UNEEG). Data-rate need: kbps aggregate;
The long-term ambulatory monitoring (no real-time decoding requirement) permits the conservative inductive-link to remain adequate.

\subsection*{Brain-spine and hybrid bidirectional loops}

This category represents the architecturally hardest systems, because they close a sensorimotor loop across two non-co-located implant sites. Onward ARC-BCI (DigitalBridge) pairs a WImagine-class cortical recorder with an ARC-IM spinal stimulator\cite{ref136,ref137}; the distinctive feature is the inter-implant coordination link, which is latency-critical (tens of milliseconds for locomotion) rather than rate-critical. CorTec's Brain Interchange BCI is implemented as a single bidirectional implant with FDA BDD in 2026\cite{ref122}. The  figures of merit for such devices are energy-per-coordination-event and worst-case loop latency, not energy-per-bit at peak rate.
%the regime where body-channel physical layers (\S2) become technically attractive.
In general, these devices are a combination of cortical recorders (epidural or subdural, depending on platform) and deep cervical or lumbar spinal cord stimulator. The two implant sites need to coordinate wirelessly, with at least one body-worn hub in the loop. Low aggregate payload per closed-loop cycle (decoded intentions and stimulation commands, kbps class) is sufficient, but worst-case end-to-end loop latency remains in the tens of milliseconds.

\subsection*{Free-floating mm-scale devices}

Free-floating implants address depths where coils are impractical, and the wireless layer is decided by the power source. Iota Biosciences' peripheral-nerve stimulators (ultrasonic power in, backscatter out\cite{ref27,ref31,ref32,ref105}) set the rate from the harvested-power budget. Motif Neurotech's magnetoelectric brain stimulator builds on MagNI\cite{ref28} and ME-BIT\cite{ref30,ref111} with FDA IDE in 2026 and skull-transparency advantages over inductive coupling\cite{ref112}; its uplink is backscatter-class because hosting a local oscillator at a few-microwatt budget remains impractical\cite{ref25,ref43,ref95}. The Berkeley/UCSF neural-dust family\cite{ref23,ref24,ref25,ref26} backscatters single-neuron data at $<$1\,nJ/b, with subsequent works demonstrating multimodal sensing plus bidirectional stimulation\cite{ref106,ref107,ref108,ref109} and multi-implant ME addressing\cite{ref33}. NIR optical powering recovers 0.1 to 1\,mW for sub-dural devices\cite{ref82,ref83,ref102,ref104,ref121}. ni2o KIWI and Subsense's magnetoelectric nanoparticles are also setting up for early-stage trials. These devices are placed in sub-mm$^3$ depths, and on peripheral nerve (Iota), sub-dural/epidural regions (Motif DOT, NIR optical), or are distributed (Subsense). There is a single intra-tissue site per implant, with external power-source coil or transducer providing both power and the uplink reference. Required data-rates range from $<$1 to a few kbps per mote in the backscatter regime; aggregate rate across a population scales with the number of addressable motes, not with per-mote link energy.

\subsection*{Established clinical neuromodulators}

The largest deployed implantable-wireless population is established neuromodulators, whose link choices set the clinical baseline. Medtronic Percept RC/PC with BrainSense replaced Activa SC/RC in 2024 and remains the canonical inductive-recharge plus sub-Mbps-MedRadio platform\cite{ref84,ref146,ref69}; Boston Scientific Vercise Genus, Abbott Infinity/Liberta RC, and Aleva directSTIM use the same template with directional leads\cite{ref84}. NeuroPace RNS pairs responsive stimulation with a magnetic-induction telemetry wand\cite{ref122}. Neuracle NEO, which is reported as the first commercially approved invasive BCI \cite{marshall2026china} contains extradural cortical electrodes, inductive recharge, conservative-class RF telemetry with data compression\cite{ref122}. Cochlear, Advanced Bionics, and MED-EL together have shipped $>$10$^5$ resonant-inductive multi-milliwatt downlinks at 1 to 13.56\,MHz\cite{ref71,ref72,ref73,ref74}.
%the reference powered-downlink physical layer.
Inspire HNS follows the same architecture\cite{ref84}, as did Argus II historically\cite{ref88,ref98}. Interestingly, the link does not have to push to sub-10\,pJ/b because the payload does not demand it
%and the evidence-base bar is institutional
\cite{ref69,ref101}. Overall, this is the category for  deep brain stimulation (STN, GPi, fornix, ANT), cortical and depth electrodes for responsive stimulation (NeuroPace RNS), cochlear implants, extradural cortex (Neuracle NEO), and hypoglossal nerve implants (Inspire). For the external device, split topology can be used, with multi-cm$^3$ pulse generator in the chest, cranial pocket (RNS), or behind-ear (cochlear). Required data rates can reach sub-Mbps aggregate for telemetry and event-driven LFP logs; the powered downlink consumes the bulk of the link energy budget, not the data uplink.

\subsection*{Insights from the commercial landscape analysis}

The platforms in Table~\ref{tab:commercial} span three orders of magnitude in volume, four in implant power, and four in aggregate rate. Five insights emerge for the wireless link from Table~\ref{tab:commercial}: (i) \emph{On-implant compression is the standard today, but not the ultimate solution}\cite{ref123,ref124,ref125,ref126,ref76}: the 100\,pJ/b to 10\,nJ/b baseline forces feature commitment, and as link energy closes toward the sub-10\,pJ/b goals, the computation may migrates off-implant, unlocking discovery beyond the fixed features.
%hand-designed compressors preserve.
(ii) \emph{Split topology dominates}: brain-adjacent electrodes plus a more energy-permissive transceiver location (chest IPG, subgaleal, behind-ear, external receiver) recurs across endovascular, brain-spine hybrid, neuromodulator, and sub-scalp categories\cite{ref8,ref9,ref12,ref84,ref122,ref146}. (iii) \emph{Modality and requirements can be dictated by surgical envelope}: intracortical implants demand gigabit-class raw uplinks, while endovascular accepts ECoG-class rates on chest radios; mm-scale free-floating motes demand omnidirectional power, with ability to operate at variable power budgets that may need configurable compression per mote. (iv) \emph{Regulatory clustering of ``firsts'' in 2024 to 2026} across distinct modalities (Precision 510(k), Epiminder De Novo, Motif IDE, INBRAIN FIH, CorTec BDD, Paradromics FIH/IDE, Kampto announcement, Neuracle approval, Medtronic Percept RC clearance) signals a taxonomy-wide commercial inflection point. (v) \emph{Wired first, wireless later} (Paradromics, Precision) decouples clinical-evidence risk from link-energy risk.
%\hl{The full 40-entry dataset is included in Supplementary Table~S1.}

%========================================================================================

\section*{4. Theoretical Comparison of Estimated Channel Capacity across Modalities}

% \subsection{Theoretical upper bounds on bi-directional channel capacity}
The preceding sections demonstrate the growing importance of wireless communication as a limiting factor in future BMIs. A diverse set of communication modalities has been analyzed, including NIC, ME, US, BCC, and UWB. Despite substantial progress, cross-modality comparisons remain challenging because reported performance depends on numerous experimental and architectural variables, including implant geometry, implantation depth, operating frequency, receiver sensitivity, and regulatory constraints.

Therefore, \emph{\ul{existing literature does not readily reveal whether the observed differences among modalities stem from fundamentally different channel characteristics or from disparities in implementation and design maturity}}. To establish a common basis for comparison, we developed a normalized channel-capacity framework encompassing the five aforementioned links. The framework assumes common operating power budget, while allowing each modality to retain its characteristic propagation losses, coupling mechanisms, and available bandwidth. Using Shannon's capacity expression, $C = B\log_{2}(1+\mathrm{SNR})$, we estimate the maximum achievable throughput for each modality and thereby compare their underlying physical limits.

\subsection{Modalities considered: Narrowband (NIC, ME, US) and Wideband (UWB, BCC)}
NIC, ME, and US employ resonant transducers optimized primarily for efficient wireless power transfer and low-bandwidth (high-$Q$) communication at relatively low carrier frequencies.
%These architectures typically achieve high coupling efficiency but comparatively narrow communication bandwidth owing to their high-$Q$ resonant behavior. 
This makes these modalities attractive for resonant downlink and wireless power transfer (WPT).
BCC and UWB occupy a contrasting design space. Rather than maximizing resonant energy transfer, these modalities exploit substantially larger bandwidths and are therefore attractive for high-rate neural-data uplinks. BCC primarily operates in the EQS regime (approximately flat band), whereas UWB exploits the wide spectral allocations, under FCC Part 15.

Together, these five modalities span the dominant physical approaches currently being explored for wireless neural links and capture both the narrowband techniques that dominate today's implant ecosystem and the broadband approaches that are increasingly becoming popular for future high-channel-count systems.

%\subsection{System model}

In the following analysis, each modality except UWB was anchored to experimentally demonstrated implant-scale systems reported in \cite{kiani_iscas25,ref144}. Transfer efficiencies, propagation losses, and transducer parameters were normalized to comparable implant dimensions and implantation depths. An equivalent receive-side representation was synthesized for all modalities to model communication link and evaluated using a common receiver noise floor. Two receiver-termination assumptions (matched and high-impedance) allowed estimation of SNR required for capacity analysis. Complete normalization procedures, source datasets, and parameter assumptions are provided in Supplementary Sections S1–S4.

\subsection{Physical limiters governing implant communication}

Although the underlying communication mechanisms differ substantially, the achievable channel capacity of all implant modalities is ultimately governed by five physical constraints:

\noindent \textbf{Tissue propagation loss}:  
Losses in the multi layer brain channel consisting of absorption, dispersion and reflections (details in Supplementary Section~S3).
    
\noindent
\textbf{Geometric and depth coupling loss}: 
Each modality exhibits a geometry-dependent coupling mechanism  and a depth depdent loss mechanism that governs link's efficiency, the trends for which are summarized in Table~\ref{tab:combined} (Part I), with complete equations in Supplementary Section~S4.     
  
\noindent 
\textbf{Transducer efficiency}: 
The conversion efficiency strongly depends on the geometry, materials, fabrication and packaging of the transducer. To avoid introducing large implementation-specific uncertainties, transducer efficiencies are extracted from experimental measurements for NIC, ME, US and BCC \cite{kiani_iscas25, ref144}. UWB's antenna efficiency is evaluated using first principles.

\noindent
\textbf{Impedance mismatch}: 
Impedance mismatch losses are considered by comparing power-matched and high impedance terminations, producing the range of expected received SNR.

\noindent
\textbf{Safety and regulatory power ceiling}: 
Regardless of channel quality, communication performance is ultimately bounded by the maximum transmit power permitted under thermal, regulatory, and biological safety constraints. These limits are modality dependent and are discussed in Supplementary Section~S4.

\subsection{Downlink capacity}

The downlink requirements of neural interfaces typically range from several kbps to at most a few Mbps, and are consequently limited by power delivery requirements rather than throughput. Despite operating in the weak-coupling regime, the SNR is large enough,
%to make the link bandwidth dominated
indicating that modalities optimized for WPT are not necessarily those that maximize communication capacity.
Thus downlink predominately use modalities optimized for power delivery like NIC, ME and US, as is evident from results shown in Table~\ref{tab:combined}.

\subsection{Uplink capacity}
In contrast to the downlink, neural-data egress places stringent demands on communication throughput. The estimated uplink capacities for the five modalities are summarized in Table~\ref{tab:combined}.
It is evident from the results that all modalities are bandwidth limited, especially the resonant links. BCC and UWB emerge as the clear choice poised to meet requirements projected for future high-channel-count neural interfaces, attributed primarily to the use of large bandwidth. The relative insensitivity of channel capacity ($C$) on SNR suggests that system and circuit level optimizations such as impedance matching and noise floor do not affect  $C$ significantly.

%\clearpage
{
{\fontsize{8pt}{5.5pt}\selectfont
\setlength{\tabcolsep}{2.2pt}
\renewcommand{\arraystretch}{0.85}
\setlength{\baselineskip}{0pt}
\setlength{\LTleft}{0pt}
\setlength{\LTright}{0pt}

\begin{longtable}
{@{}P{1.5cm}P{0.9cm}P{1.3cm}P{1.4cm}P{1.4cm}P{1.5cm}P{1.5cm}P{1.3cm}P{1.1cm}p{1.2cm}@{}}
\caption{\textbf{Geometric scaling laws and estimated channel capacity.} \textit{Part~I}: SNR, $f_r$ and $Q$ scaling laws. \textit{Part~II}: downlink channel capacity. \textit{Part~III}: uplink channel capacity.}
\label{tab:combined}\\
\toprule
\multicolumn{10}{@{}l@{}}{{\fontsize{9pt}{11pt}\selectfont \textbf{\textit{Part I: Geometric Coupling Models}}}} \\
\midrule
\textbf{Mod.} & \multicolumn{5}{c}{\textbf{$\mathrm{SNR}$ scaling with depth $d$}} & \multicolumn{2}{c}{\textbf{$f_r$ (resonant freq.)}} & \multicolumn{2}{c}{\textbf{$Q$}} \\
\midrule
NIC & \multicolumn{5}{p{7.1cm}}{$\mathrm{SNR} \propto r_1^4 r_2^4\,(r_1^2+d^2)^{-3}$} & \multicolumn{2}{p{2.95cm}}{External LC tank} & \multicolumn{2}{p{2.45cm}}{$\approx 80$ (coil)} \\
US & \multicolumn{5}{p{7.1cm}}{$\mathrm{SNR} \propto A_{\mathrm{Rx}}^2\,e^{-2\alpha_{\mathrm{ac}}d}/d^2$} & \multicolumn{2}{p{2.95cm}}{$f_r = c_s/(2h)$ (thickness)} & \multicolumn{2}{p{2.45cm}}{$\approx 5$ (fluid-loaded)} \\
ME & \multicolumn{5}{p{7.1cm}}{$\mathrm{SNR} \propto \mathrm{Vol}^2\,(r_{\mathrm{Tx}}^2+d^2)^{-3}$} & \multicolumn{2}{p{2.95cm}}{$f_{\mathrm{bar}}$ (bar resonance)} & \multicolumn{2}{p{2.45cm}}{$\approx 5.3$ (magneto-strictive)} \\
BCC & \multicolumn{5}{p{7.1cm}}{$\mathrm{SNR} \propto V_{\mathrm{tx}}^2\cdot 10^{H(d)/10}$, $H\!\propto\! d^{-2.4}$} & \multicolumn{2}{p{2.95cm}}{N/A\textsuperscript{\dag} (capacitive)} & \multicolumn{2}{p{2.45cm}}{N/A\textsuperscript{\dag}} \\
UWB & \multicolumn{5}{p{7.1cm}}{$\mathrm{SNR} \propto P_{\mathrm{tx}}\,e^{-2\alpha_{\mathrm{EM}}d}/d^2$} & \multicolumn{2}{p{2.95cm}}{N/A\textsuperscript{\ddag} (spectral mask)} & \multicolumn{2}{p{2.45cm}}{N/A\textsuperscript{\ddag} (wideband)} \\
\midrule
\multicolumn{10}{@{}p{14.5cm}@{}}{{\fontsize{6pt}{8pt}\selectfont \textsuperscript{\dag}BCC operates in the EQS quasi-static regime (DC to 100\,MHz).}}\\
\multicolumn{10}{@{}p{14.5cm}@{}}{{\fontsize{6pt}{8pt}\selectfont \textsuperscript{\ddag}UWB bandwidth and power are governed by the FCC\,\S15.517 spectral EIRP mask ($-41.3$\,dBm\,MHz$^{-1}$).}}\\
\midrule
\multicolumn{10}{@{}l@{}}{{\fontsize{9pt}{11pt}\selectfont \textbf{\textit{Part II: Downlink Shannon Channel Capacity with Tissue Corrections}}}} \\
\midrule
\textbf{Mod.} & \parbox[t]{\linewidth}{\centering\textbf{$d$}\\[2pt]{\scriptsize(mm)}} & \parbox[t]{\linewidth}{\centering\textbf{BW}\\[2pt]{\scriptsize(MHz)}} & \parbox[t]{\linewidth}{\centering\textbf{SNR}\\[2pt]{\scriptsize(dB)}} & \parbox[t]{\linewidth}{\centering\textbf{$C_{\mathrm{match}}$}\\[2pt]{\scriptsize(Mbps)}} & \parbox[t]{\linewidth}{\centering\textbf{$C_{\mathrm{High\text{-}Z}}$}\\[2pt]{\scriptsize(Mbps)}} & \multicolumn{4}{c}{\textbf{Limiting constraint}} \\
\midrule
\parbox[c]{\linewidth}{\centering NIC\\[1pt]{\fontsize{5pt}{5pt}\selectfont 33.5\,MHz}} & 10 & 0.4188 & 122.7 & 17.07 & 17.91 & \multicolumn{4}{p{5.55cm}}{SAR, FCC \S1.1310} \\
 & 20 & 0.4188 & 118.2 & 16.44 & 17.28 & \multicolumn{4}{p{5.55cm}}{} \\
 & 30 & 0.4188 & 113.0 & 15.72 & 16.56 & \multicolumn{4}{p{5.55cm}}{} \\
\midrule
\parbox[c]{\linewidth}{\centering US\\[1pt]{\fontsize{5pt}{5pt}\selectfont 238\,kHz}} & 10 & 0.0476 & 139.0 & 2.197 & 2.292 & \multicolumn{4}{p{5.55cm}}{$I_{\mathrm{SPTA}}$, FDA Track~3} \\
 & 20 & 0.0476 & 131.1 & 2.074 & 2.169 & \multicolumn{4}{p{5.55cm}}{} \\
 & 30 & 0.0476 & 128.7 & 2.035 & 2.13 & \multicolumn{4}{p{5.55cm}}{} \\
\midrule
\parbox[c]{\linewidth}{\centering ME\\[1pt]{\fontsize{5pt}{5pt}\selectfont 279.2\,kHz}} & 10 & 0.05298 & 139.2 & 2.451 & 2.557 & \multicolumn{4}{p{5.55cm}}{DRL, IEEE C95.1} \\
 & 20 & 0.05298 & 137.8 & 2.426 & 2.532 & \multicolumn{4}{p{5.55cm}}{} \\
 & 30 & 0.05298 & 135.8 & 2.389 & 2.495 & \multicolumn{4}{p{5.55cm}}{} \\
\midrule
\multicolumn{10}{@{}l@{}}{{\fontsize{9pt}{11pt}\selectfont \textbf{\textit{Part III: Uplink Shannon Channel Capacity with Tissue Corrections}}}} \\
\midrule
\textbf{Mod.} & \parbox[t]{\linewidth}{\centering\textbf{$d$}\\[2pt]{\scriptsize(mm)}} & \parbox[t]{\linewidth}{\centering\textbf{BW}\\[2pt]{\scriptsize(MHz)}} & \parbox[t]{\linewidth}{\centering\textbf{SNR}\\[2pt]{\scriptsize(dB)}} & \parbox[t]{\linewidth}{\centering\textbf{$C_{\mathrm{match}}$}\\[2pt]{\scriptsize(Gbps)}} & \parbox[t]{\linewidth}{\centering\textbf{$C_{\mathrm{High\text{-}Z}}$}\\[2pt]{\scriptsize(Gbps)}} & \parbox[t]{\linewidth}{\centering\textbf{$L_{\mathrm{tis}}$}\\[2pt]{\scriptsize(dB)}} & \parbox[t]{\linewidth}{\centering\textbf{$L_{\mathrm{path}}$}\\[2pt]{\scriptsize(dB)}} & \multicolumn{2}{c}{\parbox[t]{2.45cm}{\centering\textbf{$P_{\mathrm{tx}}$}\\[2pt]{\scriptsize(mW)}}} \\
\midrule
\parbox[c]{\linewidth}{\centering NIC\\[1pt]{\fontsize{5pt}{5pt}\selectfont 33.5\,MHz}} & 10 & 0.4188 & 104.6 & 0.01456 & 0.01539 & 0.28 & 3.07 & \multicolumn{2}{c}{3.0$^{*}$} \\
 & 20 & 0.4188 & 100.1 & 0.01393 & 0.01477 & 0.95 & 6.92 & \multicolumn{2}{c}{3.0$^{*}$} \\
 & 30 & 0.4188 & 95.0 & 0.01321 & 0.01405 & 1.31 & 11.73 & \multicolumn{2}{c}{3.0$^{*}$} \\
\midrule
\parbox[c]{\linewidth}{\centering US\\[1pt]{\fontsize{5pt}{5pt}\selectfont 238\,kHz}} & 10 & 0.0476 & 107.8 & 0.001705 & 0.0018 & 7.20 & 13.76 & \multicolumn{2}{c}{3.0$^{*}$} \\
 & 20 & 0.0476 & 100.0 & 0.001581 & 0.001677 & 7.53 & 21.25 & \multicolumn{2}{c}{3.0$^{*}$} \\
 & 30 & 0.0476 & 97.6 & 0.001543 & 0.001638 & 7.64 & 23.59 & \multicolumn{2}{c}{3.0$^{*}$} \\
\midrule
\parbox[c]{\linewidth}{\centering ME\\[1pt]{\fontsize{5pt}{5pt}\selectfont 279.2\,kHz}} & 10 & 0.05298 & 98.6 & 0.001735 & 0.001841 & 0.01 & 33.41 & \multicolumn{2}{c}{3.0$^{*}$} \\
 & 20 & 0.05298 & 97.2 & 0.00171 & 0.001816 & 0.07 & 34.78 & \multicolumn{2}{c}{3.0$^{*}$} \\
 & 30 & 0.05298 & 95.1 & 0.001673 & 0.001779 & 0.10 & 36.82 & \multicolumn{2}{c}{3.0$^{*}$} \\
\midrule
BCC & 10 & 100 & 41.8 & 1.387 & 1.587 & 0.35 & 35.67 & \multicolumn{2}{c}{3.0$^{*}$} \\
 & 20 & 100 & 32.1 & 1.066 & 1.266 & 1.16 & 44.53 & \multicolumn{2}{c}{3.0$^{*}$} \\
 & 30 & 100 & 26.4 & 0.8784 & 1.078 & 1.63 & 49.72 & \multicolumn{2}{c}{3.0$^{*}$} \\
\midrule
\mbox{UWB-700} & 10 & 700 & 47.2 & 10.98 & 12.38 & 9.03 & 4.48 & \multicolumn{2}{c}{0.4154$^{\#}$} \\
 & 20 & 700 & 41.2 & 9.582 & 10.98 & 15.32 & 10.50 & \multicolumn{2}{c}{1.7674$^{\#}$} \\
 & 30 & 700 & 31.8 & 7.405 & 8.805 & 22.65 & 14.03 & \multicolumn{2}{c}{2.49$^{+}$} \\
\mbox{UWB-B4} & 10 & 1584 & 47.2 & 24.85 & 28.02 & 9.03 & 4.48 & \multicolumn{2}{c}{0.9401$^{\#}$} \\
 & 20 & 1584 & 39.1 & 20.6 & 23.76 & 15.32 & 10.50 & \multicolumn{2}{c}{2.49$^{+}$} \\
 & 30 & 1584 & 28.3 & 14.89 & 18.06 & 22.65 & 14.03 & \multicolumn{2}{c}{2.49$^{+}$} \\
\midrule
\multicolumn{10}{@{}p{14.5cm}@{}}{{\fontsize{6pt}{7.5pt}\selectfont $^{*}$~Max.\ power limit (thermal safety, 1\,$^\circ$C budget)}}\\
\multicolumn{10}{@{}p{14.5cm}@{}}{{\fontsize{6pt}{7.5pt}\selectfont $^{\#}$~FCC~\S15.517 EIRP limit 52\,$\mu$W at 700\,MHz, 117\,$\mu$W at B4 band at the skin-air interface; $P_{tx} = P_{\mathrm{EIRP}}\times 10^{L_{\mathrm{loss}}/10}$).}}\\
\multicolumn{10}{@{}p{14.5cm}@{}}{{\fontsize{6pt}{7.5pt}\selectfont $^{+}$~SAR-1\,g limit, FCC~\S1.1310 (1.6\,W\,kg$^{-1}$ per 1\,g, grey matter).}}\\
\multicolumn{10}{@{}p{14.5cm}@{}}{{\fontsize{6pt}{7.5pt}\selectfont $^{\wedge}$~SAR-10\,g limit, ICNIRP~2020 (2.0\,W\,kg$^{-1}$ per 10\,g).}}\\
\bottomrule
\bottomrule
\end{longtable}
}
}
% \baselineskip=14pt
%\clearpage

\subsection{Impact of miniaturization}

The current analysis is anchored to implant dimensions representative of contemporary neural interfaces ($\sim$20\.mm$^{3}$). Future systems targeting higher depth are expected to shrink substantially ($\sim$1\,mm$^{3}$), motivating the question of how channel capacities change under aggressive miniaturization. Since the geometric scaling laws governing SNR, bandwidth, operating frequency and channel losses differ significantly across modalities, a crossover amongst the modalities in terms of channel capacity is possible. More in-depth analysis is captured in Supplementary Section S6.

\subsection{Implications for future neural interfaces}
\begin{enumerate}
    \item \textbf{Low-frequency, high-\emph{Q} modalities favor downlink.} By trading bandwidth for coupling efficiency and received SNR, modalities such as NIC, US, and ME allow kbps-Mbps data-rate with high PTE ($\sim$60 dB more SNR than UWB/BCC).
    %Table ~\ref{tab:combined} \emph{Part III} shows that SNR achieved by low-frequency, resonant high \emph{Q} modalities like NIC, US and ME achieve $\approx 60$ dB more power than their wide-band counterparts (BCC and UWB) under similar transmit power.

    \item \textbf{Bandwidth dominates uplink capacity for shallow-to-medium implants.} For depths below $\approx 30$ mm, SNR is already decent, and channel capacity is predominantly bandwidth-limited. Consequently, modalities offering the largest bandwidth achieve the highest uplink capacities (UWB/BCC has 100-1000$\times$ channel-capacity than NIC).
    %Table ~\ref{tab:combined} \emph{Part III} highlights the $\approx 10-20 \times$ gap in channel capacity of UWB when compared with BCC, which in turn is $\approx 100 \times$ more than NIC. %The ratio of channel capacity closely tracks the ratio of bandwidth, showcasing the strong dependence of bandwidth on capacity, rather than SNR which differs by $\approx ~60\,dB$ in the opposite sense.
    
    \item \textbf{Implant depth shifts the optimal bandwidth–loss operating point.} As implant depth increases,
    %beyond $\approx 30$ mm,
    decreasing SNR reduces the relative benefit of additional bandwidth. Modality selection therefore becomes a depth-dependent optimization between bandwidth and channel loss rather than a simple preference for the widest-bandwidth link.

    \item \textbf{Downlink and uplink may favor different operating regimes.} PTE benefits from low-frequency, resonant high-\emph{Q} operation, whereas high-rate telemetry benefits from large bandwidth. Optimal uplink and downlink modalities may differ even for the same implant.

    \item \textbf{Power and regulatory ceilings, not thermal limits, constrain wide-band uplinks.} Regulatory constraints on channel use, maximum transmit power and maximum allowable fields pose regulatory limits on the channel-capacity of broad-band modalities like UWB, well before engineering or physics based bounds.
    
    \item \textbf{Miniaturization imposes modality-specific capacity limits.} Implant scaling affects bandwidth and SNR differently across modalities. UWB is constrained by both antenna efficiency and bandwidth limits associated with small implants, whereas BCC links primarily experience reduced coupling/SNR with little change in available bandwidth.

\end{enumerate}        
    
%Future high-channel-count neural interfaces may increasingly favor uplink modalities occupying larger spectral resources for high throughput, while resonant narrow-band modalities will still be preferred for downlink PTE. Improvements in transducer design, coupling efficiency, and receiver sensitivity remain important for the downlink, but the largest gains in uplink capacity are likely to arise from modalities whose available bandwidth scales with data-rate demands.

% \input{channel_cap_analysis}

%========================================================================================
\section*{5. Future Research Directions}

The trends collected in this Review mark out a set of coupled research fronts whose advances will compound across one another. 

\begin{itemize}

\item \textbf{Towards Sub-pJ$/$b wireless links:} Today's implant transmitters operate at 100\,pJ$/$b to 10\,nJ$/$b. Reaching sub-1\,pJ$/$b requires advances on three coordinated axes: higher effective channel usage through wideband techniques, modulation/coding that approach Shannon capacity,
%in the weakly coupled regime; and %duty-cycled,
and event-driven physical-layer architectures that amortize overhead.

\item \textbf{Co-design across electromagnetics, transducers, communication theory, and circuits:} The implant transducer, tissue channel, modulation, and analog/digital front-ends sit on a common Pareto surface, with the optimum only possible through co-design, covering all four layers within a single optimization, and on open, standardized benchmarks and comparisons for PTE, energy-per-bit, channel capacity on common tissue models.

\item \textbf{On-implant compression and neural inference:} Lossy spike sorting, feature extraction, and learned codecs reach $10^{2}$ to $10^{4}\times$ data reduction, versus  $\sim$10$\times$ lossless compression. The frontier is on-implant inference via reconfigurable accelerators, with a raw-passthrough mode for discovery science, deep insights and cloud-updated model weights.

\item \textbf{Towards 10 to 50\,pJ$/$b full-SoC efficiency:} Whole-system energy must move from today's $\sim$500\,pJ$/$b (BISC: 52\,mW SoC for 108\,Mbps uplink) toward 10 to 50\,pJ$/$b through tight system-circuit co-design. Both digitized robust sensing-and-link pipelines and asymmetric ADC-less nodes that offload digitization (e.g.\ analog-to-time pulse-width modulation\cite{TD_BPQBC}) hold promise, depending on channel count, dynamic range, loss, and bandwidth.

\item \textbf{Higher-PTE links for deep implants:} PTE falls steeply with depth and transducer size, leaving 30 to 50\,mm targets (deep-brain nuclei, brainstem, spinal cord) at the hardest corner. Progress needs modality-aware downlink choice (mid-field RF, focused ultrasound, magnetoelectric, hybrids), adaptive focusing, and high-$Q$ matching within thermal limits.

\item \textbf{Privacy, security, and regulatory pathways:} The link needs physical-layer privacy, and security from near-field decay and channel directionality, post-quantum and streaming crypto primitives within sub-10\,pJ$/$b budgets, and regulatory pathways for hermetic packaging, secure firmware update, and graceful failure of chronic implants.

\item \textbf{Networks of implants:} Brain-wide closed-loop control, across shallow cortical and deep subcortical sites, exceeds any single implant. How to architect such networks, share the weakly coupled channel, and respect per-node energy and PTE limits remains an open research direction with broad clinical reach.
\end{itemize}

Across all these fronts, an efficient wireless link (powering as well as communication) holds the key to implantable BCIs with ultra-low-power, high-density recording and low-latency neuromodulation, which promises broad social impact through restoration of movement and communication, treatment of neurological and psychiatric disease, and, in time, augmentative interfaces for sensing, memory, and human-machine interaction.

\section*{Acknowledgements}
%This work was supported in part by the U.S. National Science Foundation, the Defense Advanced Research Projects Agency (DARPA), ARPA-H, the U.S. Office of Naval Research, and Purdue University.
\vspace{-3mm}
This work is supported by Quasistatics Inc. We thank members of the Sen and Chatterjee and other collaborating laboratories for stimulating discussions.

\vspace{-5mm}
\section*{Author contributions}
\vspace{-3mm}
S.S. conceived the structure of the Review. S.S., B.C., G.B., and A.R. researched data for the article, contributed to the discussion of content, wrote the article, and reviewed/edited the manuscript before submission.

\vspace{-5mm}
\section*{Competing interests}
\vspace{-3mm}
S.S. is founder of and holds equity in Quasistatics Inc., dba Ixana, which is commercializing electro-quasistatic human body communication technology. B.C. has consulted for Ixana, while A.R. has previously worked for Ixana. G.B. declares no competing interests.

\vspace{-5mm}
\section*{Use of generative AI}
\vspace{-3mm}
The authors used generative AI tools (large language models) during the preparation of this manuscript for AI-assisted research, including literature exploration and summarisation of prior work, and for copy editing of human-written text, including improvements to readability, grammar, spelling, punctuation, wording, and formatting. All scientific content, claims, citations, and figures were conceived and verified by the human authors, who take full accountability for the manuscript.
%========================================================================================

% =====================  REFERENCES  =====================
\clearpage

%% option 1 - harder to modify
% \section*{References}
% \renewcommand{\section}[2]{}  % suppress duplicate header from thebibliography
% \input{do_not_use/bibitem_bibliography}

%% option 2 - easier to modify
\bibliographystyle{unsrtnat_custom}
\bibliography{9_references_do_not_add,9_references_add_here}

\end{document}